\documentclass{jfm}
\usepackage{graphicx}
\graphicspath{{figs/}}
\usepackage{tikz}
\usepackage{amsmath, epstopdf, epsfig, multirow, float}
\usepackage{dcolumn}
\usepackage{bm}
\usepackage{xcolor}
\usepackage[hidelinks]{hyperref}
\usepackage[noabbrev]{cleveref}
\hypersetup{
  colorlinks = true,   
  urlcolor   = blue,   
  linkcolor  = blue,   
  citecolor  = black   
}

\usepackage{caption}
\usepackage{subcaption}

\usepackage[percent]{overpic}
\newcommand{\insetfig}[3][23,88]{%
    \begin{overpic}[width=\linewidth]{#2}
        \put(#1){\setlength{\fboxsep}{2pt}\colorbox{white}{\strut (#3)}}
    \end{overpic}%
}

\usepackage{multirow}
\newcommand{\cid}[1]{\texttt{#1}}

\shorttitle{Effect of initial Rayleigh mode on drop deformation}
\shortauthor{Parik, Dighe, Truscott, and Dutta}

\title{Effect of initial Rayleigh mode on drop deformation under impulsive acceleration}

\author{Aditya Parik\aff{1}
  \corresp{\email{aditya.parik@usu.edu}},
  Sandip Dighe\aff{2},
  Tadd Truscott\aff{2},
 \and Som Dutta\aff{1} \corresp{\email{som.dutta@usu.edu}}}

\affiliation{\aff{1}Department of Mechanical and Aerospace Engineering,
                 \\ Utah State University,  UT 84321, USA 
                 \aff{2}Department of Mechanical Engineering,
                 \\ King Abdullah University of Science and Technology, Saudi Arabia}

\newcommand{\Vol}{\mathcal{V}}

\newcommand{\twoD}{2\bm{D}}

\newcommand{\Ttwo}{T_\mathrm{2}}
\newcommand{\Tthree}{T_\mathrm{3}}
\newcommand{\Tfour}{T_\mathrm{4}}
\newcommand{\Tf}{T_\mathrm{f}}

\renewcommand{\Re}{\mathit{Re}}

\newcommand{\We}{\mathit{W\mkern-5mu e}}

\newcommand{\Wecr}{\We_{\mathrm{cr}}}

\newcommand{\Ohd}{\mathit{Oh}_\mathrm{d}}
\newcommand{\Oho}{\mathit{Oh}_\mathrm{o}}

\newcommand{\rhod}{\rho_\mathrm{d}}
\newcommand{\rhoo}{\rho_\mathrm{o}}
\newcommand{\mud}{\mu_\mathrm{d}}
\newcommand{\muo}{\mu_\mathrm{o}}

\newcommand{\Etot}{E}
\newcommand{\Ktot}{K}

\newcommand{\Epot}{E_{\mathrm{pot}}}

\newcommand{\Emu}{E_{\mathrm{\mu}}}

\newcommand{\Kcm}{K_{\mathrm{cm}}}

\newcommand{\Eosc}{\widetilde{E}}

\newcommand{\Kosc}{\widetilde{K}}

\newcommand{\Koscr}{\widetilde{K}_r} 
\newcommand{\Koscz}{\widetilde{K}_z} 

\newcommand{\Vtot}{\bm{V}}

\newcommand{\Vcm}{V_{\mathrm{cm,z}}}

\newcommand{\Vosc}{\widetilde{\bm{V}}}

\newcommand{\Voscr}{\widetilde{V}_r}  
\newcommand{\Voscz}{\widetilde{V}_z}

\newcommand{\ResP}{\mathcal{R}_\mathrm{P}}
\newcommand{\ResE}{\mathcal{R}_\mathrm{\Etot}}
\newcommand{\ResCM}{\mathcal{R}_\mathrm{cm}}

\begin{document}

\maketitle

\begin{abstract}
One of the fundamental ways of representing a droplet shape is through its Rayleigh-modes, where each mode corresponds to distinct surface-energy.
Previous studies have focused on the effect of these modes on free oscillations of drops.
In this paper, we systematically quantify how the different prescribed initial axisymmetric Rayleigh modes modulate aerodynamic energy uptake and the resulting deformation of an impulsively accelerated drop.
Using experimentally validated VOF-based multiphase numerical simulations, we isolate the coupled effects of finite-amplitude surface oscillation modes and the associated initial surface-energy state by initialising the drops with well-defined $(n,0)$ modes and phases $\{0,\pi\}$, while conserving the equivalent drop volume. 
We find that the deformation outcome is governed by the drag due to the drop's initial geometry, and the dynamic coupling between the free modal oscillations and the forced aerodynamic deformation. We find that constructive superposition amplifies deformation, whereas destructive superposition can stabilise the drop even when the aerodynamic forcing is sufficient to deform an analogous spherical drop to breakup.
Initial modes and phases that channel a larger fraction of the input power into deformation, in the form of oscillatory kinetic energy and additional surface energy, attain larger deformations and are closer to the fragmentation threshold.
These couplings are especially pronounced in high-viscosity systems, where viscous dissipation is large and facilitates the transfer of a larger fraction of the total energy to translational kinetic energy instead of oscillatory kinetic energy.
For low density-ratio systems, early-time coupling and energy transfer are the dominant mechanisms that govern drop deformation.
\end{abstract}


\section{Introduction}
\label{sec:introduction}

Drop deformation and breakup under impulsive acceleration are fundamental phenomena that are observed  in natural and engineered systems, e.g. raindrop breakup in clouds, liquid atomisation in engines, fragmentation of respiratory drops \citep{fabregat2021direct}, dropping fire-retardants from airtankers \citep{legendreFluidDynamicsAirtanker2024}, etc. Understanding the underlying physics of these phenomena is crucial for predicting and controlling these applications.
Dimensional analysis shows that a spherical fluid drop of density $\rhod$, viscosity $\mud$ and diameter $D$, when impulsively accelerated by an ambient fluid of density $\rhoo$, viscosity $\muo$ and velocity $V_0$, is governed by the following non-dimensional numbers $\{ \rho, \Oho, \Ohd, \We_0 \}$ \citep{Hinze1955, guildenbecherSecondaryAtomization2009}.
The initial Weber number $\We_0 = \rhoo V_0^2 D/\sigma$ represents the ratio of disruptive aerodynamic pressure forces to restorative surface tension forces. The density ratio, $\rho = \rhod/\rhoo$, is the ratio of the density of the drop to that of the ambient fluid.
A higher density ratio represents the drop to have a larger inertia and thus lower centre-of-mass accelerations for equivalent aerodynamic forcing.
The drop Ohnesorge number $\Ohd$ $(= \mud/\sqrt{\rhod\,\sigma\,D})$ represents the ratio of the timescale over which momentum diffuses due to viscosity, to the timescale for a capillary wave to traverse the length of the drop.
A small $\Ohd$ thus represents a drop where local surface perturbations maintain their locality longer, without dissipating over the entire drop.
The ambient Ohnesorge number $\Oho$ $(= \muo/\sqrt{\rhoo\,\sigma\,D})$ is an analogous number for the ambient fluid, and provides an alternative to Reynolds number (inversely proportional to $\Re$) of the flow around the drop that solely depends on the physical properties of the ambient fluid.
Whether a drop will deform enough under the action of ambient aerodynamic forces overcoming the restorative surface tension forces and viscous dissipation to eventually fragment is governed by a single $\Pi$-group that combines these four non-dimensional parameters.
The minimum external forcing at which a drop fragments is often referred to as the fragmentation threshold, and $\Pi$ should predict when a drop--external-flow combination meets this criterion.
\cite{Hinze1955} first proposed that $\Pi$ could be represented by the initial Weber number ($\We_0$) of the drop.
This concept was refined by \cite{Hsiang1995}, who showed that the critical Weber number for fragmentation increases with drop viscosity: more viscous drops ($\Ohd\gg 1$) resist fragmentation until a higher $\We_0$ value is reached.
Analytical models predicting this threshold, expressed as the critical Weber number $\Wecr$, have also been developed.
\cite{Villermaux2009} produced an analytical relationship for $\Wecr$ for inviscid high density-ratio systems ($\rho>500$), i.e., drop-ambient systems that form a flat pancake and show negligible centre-of-mass accelerations.
Subsequent works improved this analytical relationship to account for a non-zero drop viscosity \citep{Kulkarni2014}, again limited to $\rho>500$.
However, when the parameter space was extended to a wider range of $\rho$, $\Oho$, and  $\Ohd$ past the previously explored limits, $\Wecr$ was found to be a poor predictor of the threshold of drop breakup \citep{parikThresholdDropFragmentation2025}.
A new non-dimensional number $C_\mathrm{breakup}$, which represents the competition between the forces that drive drop deformation and the forces that resist drop deformation, was derived by \cite{parikThresholdDropFragmentation2025}.
$C_\mathrm{breakup}$ captures the influence of all the relevant non-dimensional parameters ($\rho$, $\Oho$, $\Ohd$) and contextualises $\Wecr$ as an independent parameter that affects this breakup criterion.

However, a critical assumption in this functional relationship is that the drop is initially perfectly spherical, which implies that the drop is in the state of minimum surface potential energy.
This is rarely valid in practice.
For example, raindrops almost always show oscillations with an amplitude closely matching their equilibrium terminal velocity shapes, with the $(2,0)$ Rayleigh mode being the most common \citep{Szakall2009}.
Drops in many real-world applications are typically products of a primary fragmentation (e.g., from a jet or sheet), which subsequently undergo secondary fragmentation.
This formation process imparts a non-spherical initial shape and may even have a complex internal velocity field \citep{shinjoSimulationLiquidJet2010,pairettiMeshResolutionEffects2020}. Consequently, secondary drops begin not in a static spherical state but in a dynamic state of shape oscillations, which can be represented as a superposition of multiple oscillation modes.
The first study of drop shape oscillations was that of \cite{rayleighVICapillaryPhenomena1879}, who showed that the natural oscillation modes of an inviscid drop's surface are spherical harmonics, henceforth referred to as Rayleigh modes.
Each Rayleigh mode is characterised by mode numbers $(n,m)$, with a characteristic frequency $\omega_n$ dependent on surface tension $\sigma$, density $\rhod$, and drop diameter $D$.
\begin{equation}
  \label{eq:Rayleigh_frequency}
  \omega_n^2 = \dfrac{8 n(n-1)(n+2) \sigma}{\rhod D^3}.
\end{equation}
The corresponding radial function that describes the distance of the drop interface from the drop centre, $r_{n,m}$, for a given mode $(n,m)$, non-dimensional amplitude $\tilde A_{n,m}$ ($=A_{n,m}/D$), and phase $\psi = \omega_{n,m}t + \psi_0$, can be described using associated Legendre functions $P_{n,m}$ as
\begin{equation}
  \label{eq:rayleigh_shape_nm}
  r_{n,m}(t,\theta,\phi) = D \left[ 0.5 + \tilde A_{n,m} \cos(\psi)\;P_{n,m}(\cos\theta)\;\cos(m\phi) \right].
\end{equation}
\begin{figure}
    \centering
    \includegraphics[width=0.6\linewidth]{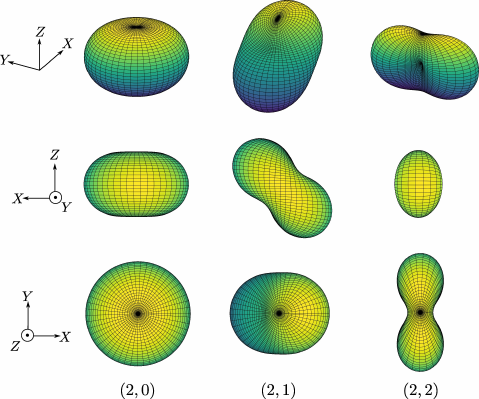}
    \caption{The harmonic modes for the fundamental oscillation $n=2$ at an extreme phase. The $(2,0)$ ``zonal'' mode is axisymmetric, while the $(2,1)$ ``tesseral'' and $(2,2)$ ``sectoral'' modes are non-axisymmetric.}
    \label{fig:degeneratemodes}
\end{figure}

For each fundamental mode $n$, there exist $0\le m \le n$ unique auxiliary or harmonic modes that display differing orientations.
For a drop undergoing free oscillations of small amplitudes, these modes are degenerate, i.e., the oscillation frequency is exactly the same for all $m$.
However, this degeneracy is broken for highly deformed drops or those under the action of external forces.
Figure \ref{fig:degeneratemodes} shows the three harmonic modes for $n=2$.
The $m=0$ mode is axisymmetric and is also described as a ``zonal'' mode, resulting from a stretching or flattening along the $z$-axis.
The $m=1$ and $m=2$ modes, on the other hand, are non-axisymmetric.
The $m=1$ mode (``tesseral'') results from a twist or shear, while the $m=2$ mode (``sectoral'') produces $4$ lobes deforming into the equatorial plane.
This concept of the $(n,0)$ mode being ``zonal'', the $(n,n)$ mode being ``sectoral'', and all other harmonics showing ``tesseral'' geometry exists for all fundamental modes.
\begin{figure}
    \centering
    \includegraphics[width=1.0\linewidth]{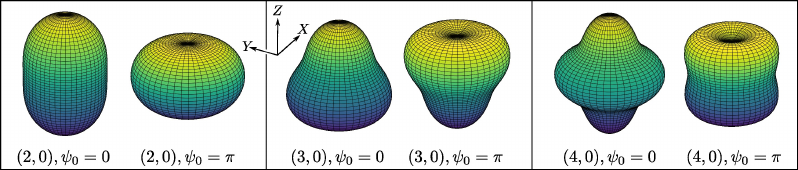}
    \caption{The figure shows 3-dimensional renders of the two extreme deformation phases, $0$ and $\pi$, for the $(2,0)$, $(3,0)$, and $(4,0)$ axisymmetric Rayleigh oscillation modes, which are the focus of the current study.}
    \label{fig:m0_modes}
\end{figure}
In the current study, we focus exclusively on the axisymmetric ``zonal'' modes ($m=0$), as they represent a primary deformation relevant to the axisymmetric forcing. The specific modes under investigation, $(2,0)$, $(3,0)$, and $(4,0)$, are illustrated in Figure \ref{fig:m0_modes}.

Subsequent work on drop oscillations revealed a more complex picture.
Viscosity, while classically known to damp oscillations \citep{lambHydrodynamics1932,chandrasekharOscillationsViscousLiquid1959}, was found to have a profound and non-trivial impact on the dynamics.
Viscosity strongly influences coupling between modes, causing higher modes to damp more quickly \citep{meradjiNumericalSimulationLiquid2001} and significantly enhancing resonant interactions between modes \citep{basaranNonlinearOscillationsViscous1992,beckerNonlinearDynamicsViscous1994}.
Furthermore, for amplitudes exceeding $\sim$10\% of the drop radius, nonlinear effects dominate \citep{beckerExperimentalTheoreticalInvestigation1991}, including a characteristic decrease in oscillation frequency with increasing amplitude \citep{brantfooteNumericalMethodStudying1973,trinhLargeamplitudeFreeDriven1982} and an asymmetry in the oscillation period \citep{trinhLargeamplitudeFreeDriven1982,tsamopoulosNonlinearOscillationsInviscid1983}.
Recently, \cite{zrnicWeaklyNonlinearShape2022} found that even a second-order nonlinear approximation captures the frequency decrease in viscous drops, in contrast to the inviscid case \citep{zrnicWeaklyNonlinearShape2021}. Viscosity has been shown to further enhance mode coupling, preferentially transferring energy from higher modes to the fundamental mode $n=2$.
The kinetic energy stored in the internal velocity field imparted during formation can be as significant as the initial surface energy \citep{zhangShorttermOscillationFalling2019}.
This initial internal circulation can amplify the oscillation amplitudes, induce phase shifts, and cause significant energy transfer between modes.
Thus, a ``non-spherical'' drop is a complex dynamic system with a pre-existing energy state that dictates its subsequent evolution.
Despite this rich understanding of free oscillations, research on initially deformed drops under external forcing is scarce, and has focused primarily on drops falling under gravity \citep{Szakall2009,szakallShapesOscillationsFalling2010,agrawalNonsphericalLiquidDroplet2017,agrawalExperimentalInvestigationNonspherical2020,ballaShapeOscillationsNonspherical2019}.
To our knowledge, no systematic study has explored how a drop with a well-defined initial oscillation mode behaves under an instantaneous force loading, characteristic of impulsive acceleration.

On the other hand, various analytical studies have attempted to describe the deformation of an initially spherical drop under impulsive acceleration \citep{Villermaux2009,Kulkarni2014,jackiwAerodynamicDropletBreakup2021}, including inviscid and viscous extensions.
One such work that provides a particularly useful framework for the current study is that of \citet{rimbertSpheroidalDropletDeformation2020}.
In their analytical model, the deformation expressed by a drop under impulsive acceleration is restricted to spheroidal shapes, such that a single degree of freedom describes the entire deformation path.
An energy balance between the kinetic energy of deformation, surface-energy variation, viscous dissipation, and the work done by the external medium through pressure forces then yields an evolution equation for the droplet shape.
Solving this evolution equation provides an analytical description of the deformation (and, in their framework, the transition between oscillations about a deformed equilibrium and unstable growth).
Importantly, this explicit energetics (power) viewpoint also allows the model to be initialised from non-spherical (prolate or oblate) spheroidal states, which can be interpreted as drops with stored surface energy at $t=0$.
Since a non-spherical initial shape directly affects both the initial aerodynamics and the initial surface-energy content of the drop, an energetics viewpoint provides an intuitive way to quantify how the external forcing is subsequently partitioned between recoverable oscillatory (deformation) energy and viscous losses.
This perspective is useful for our work, where we similarly leverage an energy/power budget to diagnose how prescribed initial shape content modifies the coupling between the forced deformation and the drop's internal response, and ultimately modulates the resulting deformation and the drop's susceptibility to breakup.

This suggests that differences in deformation outcome may be traced to how efficiently the external forcing transfers energy into deformation.
When the aerodynamic forcing couples strongly and constructively with any pre-existing oscillation imposed on the drop-ambient interface, a large fraction of input power can be channelled into internal flow, effectively amplifying radial expansion.
In particular, the coupling between aerodynamic forcing and drop shape oscillations can be significant when the aerodynamic deformation timescale ($\tau_{\mathrm{D}}$) and the Rayleigh oscillation period ($2\pi/\omega_n$) are comparable, especially for the fundamental $n=2$ mode.
For an integer $N$, this can be expressed as
\begin{equation}
  \tau = \frac{D\sqrt{\rho}}{V_0} \sim N\frac{2\pi}{\omega_n}.
  \label{eq:deformation_timescale}
\end{equation}
This temporal proximity suggests that the ``natural'' oscillations from the initial shape can appreciably interact with the ``forced'' deformation from the aerodynamic load.
This interaction could be constructive, amplifying deformation, or destructive, thus stabilising the drop.
The drop's initial energy state thus alters the deformation trajectory, and with it the susceptibility to fragmentation.
We expect drop viscosity ($\Ohd$) to be a key mediator, damping the initial oscillations and thus controlling the magnitude of this interaction.

In the current paper, we quantify the effect of this initial axisymmetric modal shape and its associated energy on the drop's subsequent deformation, internal flows, and overall energetics under impulsive acceleration.
We use the modal definition of \cite{rayleighVICapillaryPhenomena1879} to systematically define initial shapes and use VOF-based direct numerical simulations to analyse the resulting drop dynamics.

\section{Problem description and model formulation}
\label{sec:problem_formulation}

\subsection{Problem definition}
\label{subsec:physical_problem_description}
As described in section \ref{sec:introduction}, a drop-ambient system under impulsive acceleration is completely characterised by the non-dimensional set $\{ \rho, \Oho, \Ohd, \We_0 \}$.
\cite{parikThresholdDropFragmentation2025} found that the threshold is strongly dependent on all three parameters, with $\Wecr$ achieving values from $10$ (for water-air like systems) to $70$ (for $\rho\sim 10$ systems). Non-trivial morphologies are also observed, such as forward pancakes for low density low Reynolds number systems, and large drop viscosity is found to suppress plume formation due to its high viscous dissipation.

In order to efficiently study the effect of the initial drop shape on drop deformation and breakup, we consider a representative subset of $\{ \rho, \Oho, \Ohd, \We \}$ that will capture the wide variety of deformation characteristics observed in \cite{parikThresholdDropFragmentation2025}.
More specifically, the following three impulsively accelerated representative cases are chosen to represent the range of possible deformation characteristics:
\begin{enumerate}
  \item A water drop exposed to a jet of ambient air, with the $\We_0$ high enough that it will result in backward bag breakup for a spherical drop.
  The chosen parameters represent one of the most commonly studied scenarios for drop deformation and breakup.
  \item A Newtonian liquid that is $100$ times more viscous than water, exposed to a jet of ambient air. A representative case for systems such as retardant fluids used in aerial firefighting.
  \item A water drop impulsively accelerated in another liquid with densities and viscosities in the same order of magnitude, representative of a liquid-liquid system.
\end{enumerate}

A water drop impulsively accelerated in air at threshold conditions is defined by the following non-dimensional parameters: $\rho = 815.9$, $\Ohd = 1.26\times 10^{-3}$, $\Oho = 6.96\times 10^{-4}$, and $\We = 12$.
The non-dimensional numbers for the other two drop-ambient systems will be defined relative to the water-air system.
For the three cases listed above, we will impose three different initial axisymmetric oscillation modes on the drop: $(2,0)$, $(3,0)$, and $(4,0)$, using the general Rayleigh description of sinusoidal surface oscillations described by equation \ref{eq:rayleigh_shape_nm}.
Rather than the commonly used linearised simplification, we retain Rayleigh's original volume constraint so that every imposed shape encloses the same volume as the target sphere; this volume-consistent form is given by equation \ref{eq:rayleigh_shape_n0_general_main} used with equation \ref{eq:volume_conservation_final_main}, and will be described in section \ref{subsec:rayleigh_modes_corrected}.
For each mode, two different initial phases $\psi_\mathrm{0}$ will be considered such that the influence of starting at any of the two extreme states of the oscillation mode is studied.
The prolate shape observed in the experiments in section~\ref{subsec:dropshape_exp_comparison} can be approximated as a $(2,0)$ Rayleigh mode oscillation with an amplitude of $\tilde A_n \approx 0.15$, and is set as the oscillation amplitude for all the simulations comprising the parametric sweep.
Figure \ref{fig:problem_description}(b) shows an illustration (to scale) of the three modes and the two initial phases of amplitude $\tilde A_n = 0.15$ imposed on a sphere of diameter $1$.

We deliberately hold two parameters fixed across the entire parameter sweep: the oscillation amplitude $\tilde A_n = 0.15$ and the Weber number $\We_0$, which for each system is set at (or marginally above) the fragmentation threshold $\Wecr$ of its spherical reference drop.
The amplitude is chosen to match the finite-amplitude prolate shapes observed experimentally (section~\ref{subsec:dropshape_exp_comparison}). $\tilde A_n = 0.15$ is large enough for weakly nonlinear modal coupling to be important \citep{beckerExperimentalTheoreticalInvestigation1991}, and yet small enough that the resulting oscillation frequencies can still be estimated from linear Rayleigh theory to leading order \citep{trinhLargeamplitudeFreeDriven1982,zrnicWeaklyNonlinearShape2022}.
This allows us to safely utilise Rayleigh oscillation frequencies for discussing the early-stage coupling between ambient forces and initial modes.

The Weber number $\We_0$ is fixed near threshold because this is the regime in which the initial mode is most consequential: the drop sits squarely at the stability limit, and even relatively small changes in the energy balance produced by the constructive or destructive coupling of the initial shape with aerodynamic forces can tip the drop's fate from fragmentation to survival, and vice versa.
The influence of initial shape is expected to wane progressively with increase in $\We_0$, becoming negligible in the most extreme, \emph{catastrophic} regime \citep{raoSecondaryAtomizationDroplets2026}, where rapid Rayleigh--Taylor piercing \citep{theofanousAerobreakupNewtonianViscoelastic2011} penetrates the entire liquid mass almost instantaneously (relative to $\tau$) and erases the memory of any imposed global mode.
More fundamentally, our findings are most representative of breakup that develops along the deformation pathway explored in this work: global flattening into a pancake, progressive thinning of that pancake into a bag, and fragmentation on the order of the aerodynamic deformation timescale.
The influence of initial drop shape on higher $\We_0$ systems is briefly touched upon in section~\ref{sec:dropshape_conclusions}.

All the cases considered in this work are summarised in the table \ref{tab:simparams} with the final column specifying a \texttt{case ID} for every case.
We will use the \texttt{case ID}s to easily reference the cases in all subsequent text.
\begin{table}
\centering
\begin{tabular}{c|cccccccc}
\begin{tabular}[c]{@{}c@{}}Drop-Ambient\\ System\end{tabular}                           & $\rho$   & $\Oho$              & $\Ohd$              & $\We$ & $\tilde A_n$ & $n$ & $\psi_\mathrm{0}$ & Case \texttt{ID} \\ \hline
\multirow{6}{*}{\begin{tabular}[c]{@{}c@{}}Water Drop\\ in Air\end{tabular}}            & $815.9$ & $6.96\times10^{-4}$ & $1.26\times10^{-3}$ & $12$  & $0.15$ & $2$ & $0$    & \cid{WA20}       \\
                                                                                        & $815.9$ & $6.96\times10^{-4}$ & $1.26\times10^{-3}$ & $12$  & $0.15$ & $2$ & $\pi$  & \cid{WA2P}       \\
                                                                                        & $815.9$ & $6.96\times10^{-4}$ & $1.26\times10^{-3}$ & $12$  & $0.15$ & $3$ & $0$    & \cid{WA30}       \\
                                                                                        & $815.9$ & $6.96\times10^{-4}$ & $1.26\times10^{-3}$ & $12$  & $0.15$ & $3$ & $\pi$  & \cid{WA3P}       \\
                                                                                        & $815.9$ & $6.96\times10^{-4}$ & $1.26\times10^{-3}$ & $12$  & $0.15$ & $4$ & $0$    & \cid{WA40}       \\
                                                                                        & $815.9$ & $6.96\times10^{-4}$ & $1.26\times10^{-3}$ & $12$  & $0.15$ & $4$ & $\pi$  & \cid{WA4P}       \\ \hline
\multirow{6}{*}{\begin{tabular}[c]{@{}c@{}}High Viscosity\\ Drop in Air\end{tabular}}   & $815.9$ & $6.96\times10^{-4}$ & $1.26\times10^{-1}$ & $18$  & $0.15$ & $2$ & $0$    & \cid{HA20}       \\
                                                                                        & $815.9$ & $6.96\times10^{-4}$ & $1.26\times10^{-1}$ & $18$  & $0.15$ & $2$ & $\pi$  & \cid{HA2P}       \\
                                                                                        & $815.9$ & $6.96\times10^{-4}$ & $1.26\times10^{-1}$ & $18$  & $0.15$ & $3$ & $0$    & \cid{HA30}       \\
                                                                                        & $815.9$ & $6.96\times10^{-4}$ & $1.26\times10^{-1}$ & $18$  & $0.15$ & $3$ & $\pi$  & \cid{HA3P}       \\
                                                                                        & $815.9$ & $6.96\times10^{-4}$ & $1.26\times10^{-1}$ & $18$  & $0.15$ & $4$ & $0$    & \cid{HA40}       \\
                                                                                        & $815.9$ & $6.96\times10^{-4}$ & $1.26\times10^{-1}$ & $18$  & $0.15$ & $4$ & $\pi$  & \cid{HA4P}       \\ \hline
\multirow{6}{*}{\begin{tabular}[c]{@{}c@{}}Water Drop in\\ Another Liquid\end{tabular}} & $10.0$   & $1.26\times10^{-3}$ & $1.26\times10^{-3}$ & $10$  & $0.15$ & $2$ & $0$    & \cid{LL20}       \\
                                                                                        & $10.0$   & $1.26\times10^{-3}$ & $1.26\times10^{-3}$ & $10$  & $0.15$ & $2$ & $\pi$  & \cid{LL2P}       \\
                                                                                        & $10.0$   & $1.26\times10^{-3}$ & $1.26\times10^{-3}$ & $10$  & $0.15$ & $3$ & $0$    & \cid{LL30}       \\
                                                                                        & $10.0$   & $1.26\times10^{-3}$ & $1.26\times10^{-3}$ & $10$  & $0.15$ & $3$ & $\pi$  & \cid{LL3P}       \\
                                                                                        & $10.0$   & $1.26\times10^{-3}$ & $1.26\times10^{-3}$ & $10$  & $0.15$ & $4$ & $0$    & \cid{LL40}       \\
                                                                                        & $10.0$   & $1.26\times10^{-3}$ & $1.26\times10^{-3}$ & $10$  & $0.15$ & $4$ & $\pi$  & \cid{LL4P}       \\
\end{tabular}
\caption{All drop-ambient systems simulated for this study are presented in this table. A ``case \texttt{ID}" is specified for each case for each initial mode and phase for easy reference in the text.}
\label{tab:simparams}
\end{table}

\subsection{Volume consistent Rayleigh oscillation modes at finite amplitude}
\label{subsec:rayleigh_modes_corrected}
For this work, we aim to generate initial drop shapes based on axisymmetric Rayleigh modes, corresponding to the $(n,0)$ modes in the equation \ref{eq:rayleigh_shape_nm}.
At time $t=0$, the shape is a superposition of all such modes from $n=2$ to $\infty$, given as
\begin{equation}
\label{eq:rayleigh_shape_n0_simple}
\tilde r_n(\theta) = 0.5 + \sum_{n=2}^{n\to\infty} \tilde A_n P_n (\cos\theta).
\end{equation}
$\tilde A_n$ is the amplitude of the $n$\textsuperscript{th} mode, and $P_n$ is the associated Legendre polynomial, with $\tilde \square$ representing lengths non-dimensionalised by drop diameter $D$.
However, this classical formulation, derived under the assumption of small (linearised) perturbations, has a critical limitation --- applying equation~(\ref{eq:rayleigh_shape_n0_simple}) at finite amplitudes yields an initial shape whose enclosed volume, and therefore its volume-averaged diameter, exceeds the target $\tilde D = 1$.
This volume discrepancy grows with amplitude, reaching increases in volume of $\sim 7\%$ for $\tilde A_n = 0.15$ and $\sim 24\%$ for $\tilde A_n = 0.3$ for the $(2,0)$ mode, as quantified in figure~\ref{fig:corrected_rayleigh_modes} in the Appendix.
Since both the Weber number and drop inertia depend directly on this diameter, all comparable simulations must begin from the same initial volume. This makes equation~(\ref{eq:rayleigh_shape_n0_simple}) by itself unsuitable as an initial condition for the current simulations.

We address this by following Rayleigh's original treatment \citep{rayleighVICapillaryPhenomena1879}. In Appendix~II of the cited work, the drop surface is expanded as $r = a_0 + \sum_n a_n P_n(\cos\theta)$, where the base radius $a_0$ is \emph{not} identified with the equilibrium-sphere radius $a$; rather, it is determined by enforcing that the deformed drop enclose the same volume as the sphere of radius $a$, leading to equation (34) in \cite{rayleighVICapillaryPhenomena1879},
\begin{equation}
\label{eq:rayleigh_eq34}
a^3 = a_0^3\left[\,1 + 3a^{-2} \sum_n (2n+1)^{-1} a_n^2\,\right].
\end{equation}
The simplified description~(\ref{eq:rayleigh_shape_n0_simple}) corresponds to the linear limit $a_0 \to a$, in which this constraint is discarded.
Non-dimensionalising Rayleigh's shape expansion by $D$ and retaining $a_0 \neq a$ leads to
\begin{equation}
\label{eq:rayleigh_shape_n0_general_main}
\tilde r_n(\theta) = \tilde b_0 + \sum_{n=2}^{n\to\infty} \tilde A_n P_n (\cos\theta),
\end{equation}
where $\tilde b_0 = a_0/D$ is the non-dimensional counterpart of Rayleigh's base radius $a_0$ and departs from $0.5$ at finite amplitude.
As Rayleigh did, we fix $\tilde b_0$ by equating the volume enclosed by the deformed drop to that of the target sphere of radius $0.5$, integrating the shape equation over $0$ to $\pi$:
\begin{equation}
\label{eq:volume_conservation1_main}
\dfrac{\pi}{6} = \dfrac{2\pi}{3} \int_{\theta=0}^\pi [\tilde r_n(\theta)]^3\, \sin\theta\; d\theta.
\end{equation}
By substituting the shape equation \ref{eq:rayleigh_shape_n0_general_main} into this volume integral and performing the integration without invoking the linear approximation (detailed in appendix~\ref{app:derivation_b0}), we arrive at a cubic equation for the non-dimensionalised base radius, $\tilde b_0$:
\begin{equation}
\label{eq:volume_conservation_final_main}
1 = 8 \tilde b_0^3
+ 24 \tilde b_0 \sum_{n=2}^{n\to\infty} \dfrac{\tilde A_n^2}{2n+1}
+ 8 \sum_{i,j,k}^{\infty} \tilde A_i \tilde A_j \tilde A_k
\begin{pmatrix} i & j & k \\
0 & 0 & 0
\end{pmatrix}^2.
\end{equation}
The first two terms reproduce Rayleigh's equation~(\ref{eq:rayleigh_eq34}) in non-dimensional form; the final term is the finite-amplitude (cubic) contribution that the linearised analysis omits. By not assuming small-amplitude linearity, we thus extend Rayleigh's volume constraint into an exact relation valid for any superposition of axisymmetric modes at arbitrary amplitude.
This equation can be readily solved for $\tilde b_0$ using a standard numerical root-finding algorithm given $0<\tilde b_0 \le 0.5$. Using the obtained $\tilde b_0$ in equation \ref{eq:rayleigh_shape_n0_general_main} provides an initial drop geometry that matches the volume of the target sphere of diameter $1$ for any arbitrary superposition of axisymmetric Rayleigh modes at any amplitude.

\subsection{Governing equations and solver}
\label{subsec:solver}
\begin{figure}
\centering
\includegraphics[width=0.9\textwidth]{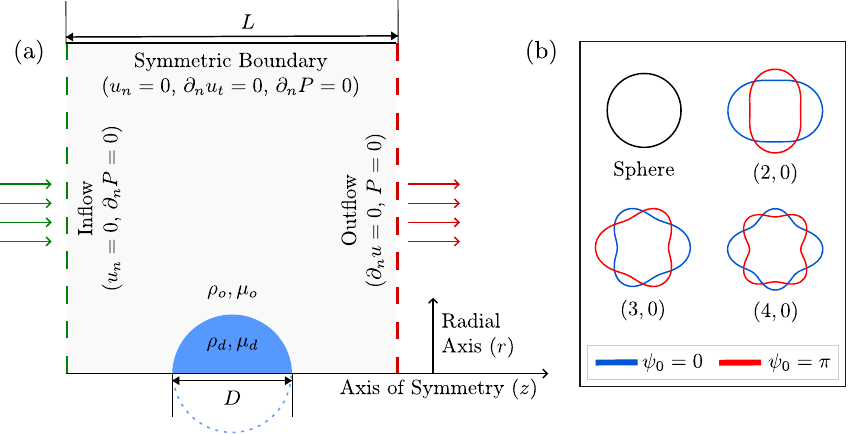}
\caption{\label{fig:problem_description}
  (a) The computational domain for the problem under consideration.
  (b) The three oscillation modes and the corresponding two initial phases of amplitude $\tilde A_n = 0.15$ imposed on a sphere of diameter $1$.}
\end{figure}

The simulations are performed using the open-source solver Basilisk \citep{popinetGerrisTreebasedAdaptive2003,Popinet2009}, which is extensively validated for two-phase flows \citep{Popinet1999,popinetGerrisTreebasedAdaptive2003,Popinet2009,popinetQuadtreeadaptiveMultigridSolver2015,vanhooftAdaptiveGridsAtmospheric2018,farsoiyaBubblemediatedTransferDilute2021}.
Basilisk solves the incompressible two-phase Navier-Stokes equations using a one-fluid formulation:
\begin{subequations}
  \label{eq:NS}
  \begin{align}
    \rho(\partial_t \bm{u} + \bm{u}\cdot\bm{\nabla u}) &= -\bm{\nabla} p + \bm{\nabla}\cdot(\mu \twoD) + \sigma \kappa \delta_s \bm{n},
    \\
    \bm{\nabla \cdot u} &= 0.
  \end{align}
\end{subequations}
where $\bm{u}$ is the fluid velocity, $\rho$ is the density, $\mu$ is the dynamic viscosity, $p$ is the pressure, and $\twoD = \bm{\nabla u} + (\bm{\nabla u})^T$ is the rate-of-deformation tensor.
Surface tension is modelled as a body force using a Continuum Surface Force (CSF) approach \citep{brackbillContinuumMethodModeling1992}. The $\delta_s$ function applies this force only at the interface, where $\sigma$ is the surface tension coefficient, $\bm{n}$ is the outward unit normal pointing from the drop into the ambient, and $\kappa$ is the signed interface curvature, taken positive where the interface is convex toward the ambient. A balanced-force formulation and height-function-based curvature estimation are employed to minimise parasitic currents \citep{francoisBalancedforceAlgorithmContinuous2006,Popinet2009}.
In the sharp-interface limit, with constant surface tension and no mass transfer, the one-fluid formulation recovers
\begin{subequations}
  \begin{align}
  [\bm{u}]_\Gamma &= 0, \\
  [\bm{n}\cdot\bm{T}\cdot\bm{n}]_\Gamma &= \sigma\kappa, \\
  [\bm{t}\cdot\bm{T}\cdot\bm{n}]_\Gamma &= 0,
  \end{align}
\end{subequations}
where $\bm{t}$ is any unit vector tangent to the interface, and $\bm{T}=-p\bm{I}+\mu\twoD$ and $[\,\cdot\,]_\Gamma$ denotes the jump in the direction of $\bm{n}$ (ambient minus drop).
Because the isotropic part of $\bm{T}$ is $-p$, this normal-stress condition is a drop in pressure on crossing from the drop interior to the ambient.
The CSF term supplies this normal (capillary) contribution.
The continuity of velocity and of tangential traction follows from the one-fluid formulation of the momentum equation.
We note that gravitational body forces are not included in the simulations performed for the parametric sweep, but are retained in the simulations utilised for the comparison with experiments (section~\ref{subsec:dropshape_exp_comparison}).

The one-fluid model defines local properties based on the volume fraction $c(\bm{x}, t)$ of one fluid (e.g., the drop, $\rhod, \mud$) relative to the other (e.g., the ambient, $\rhoo, \muo$):
\begin{subequations}
  \label{eq:vof}
  \begin{align}
    \rho &= c\,\rhod + (1-c)\,\rhoo,
    \\
    \mu &= c\,\mud + (1-c)\,\muo.
  \end{align}
\end{subequations}
The interface is tracked by advecting the volume fraction field $c$ using a geometric VOF method given as
\begin{align}
  \label{eq:c_adv}
  \partial_t c + \bm{\nabla}\cdot(c\,\bm{u}) = 0.
\end{align}
The solver utilises a quadtree-based adaptive mesh, which dynamically refines the grid at the fluid interface, enabling high-resolution capture of capillary phenomena at a reduced computational cost.
For a more detailed description of the numerical methods and implementation, we refer the reader to \cite{popinetGerrisTreebasedAdaptive2003,Popinet2009}.

\subsection{Computational setup and initial conditions}
\label{subsec:setup}
This work aims to understand how an initially imposed modal deformation modulates the subsequent aerodynamic energy transfer to the drop interface.
To track the temporal composition of this energy across interface modes, we restrict the parametric study to axisymmetric simulations, which resolve initial modes of the form $(n,m)$ with $m=0$.
The computation is carried out in an axisymmetric square domain of size $L$, as illustrated in figure \ref{fig:problem_description}(a).
The domain size $L$ is set to $16D$ for high-inertia drops ($\rho \ge 100$) and $32D$ for low-inertia drops ($\rho = 10$) to ensure that the drop remains a sufficient distance from the boundaries.
The left boundary is a uniform inflow ($u_n = V_0 = 1$), the right boundary is a free outflow ($\partial_nu_n=0$), the bottom boundary is the axis of symmetry, and the top boundary is a symmetric boundary.
At $t=0$, the drop (with a volume-averaged diameter $D=1$) is initialised with zero velocity.
Each drop is started at the extreme state of its prescribed oscillation mode ($\psi_0 = 0$ or $\pi$), where surface displacements are maximal and the oscillatory internal velocities are identically zero.
This isolates the role of the initial shape and its surface energy from that of a pre-existing internal flow.
Rest also omits the interfacial shear that such a flow would impose; the subsequent shear is set by the aerodynamic stresses.
The same turning-point state is realised in the experiments of section~\ref{subsec:dropshape_exp_comparison}, which further supports the applicability of this initial condition.
These shapes correspond to the $(2,0)$, $(3,0)$, and $(4,0)$ axisymmetric Rayleigh modes with an initial amplitude of $\tilde A_n = 0.15$, as detailed in section \ref{subsec:physical_problem_description} and illustrated in figure \ref{fig:problem_description}(b).

The solution accuracy is controlled by the wavelet error thresholds for the velocity ($\chi_u$) and volume fraction ($\chi_c$) fields, the tolerance of the Poisson solver ($\epsilon_p$), and the maximum refinement level ($N$).
Based on the detailed convergence study presented in our previous work \citep{parikThresholdDropFragmentation2025}, we use the following converged parameters for all simulations: 1) Wavelet error for velocity $\chi_u = 10^{-4}$, 2) Wavelet error for volume fraction $\chi_c = 10^{-6}$, and 3) Poisson solver residual $\epsilon_p = 10^{-4}$.
The minimum allowed cell size is set to $1024$ cells per diameter ($N=14$ for $L=16$) for all cases.
Finally, some energy and dissipation metrics are derived from the simulations and are described in section \ref{sec:discussion}.
It is essential that we test that these higher order metrics are also converged for the simulation parameters chosen for all the cases in order to ensure their veracity.
We present this test in appendix \ref{app:convergence}.
We also employ two additional numerical treatments to ensure stability and accuracy.
For high density-ratio simulations ($\rho > 500$), the volume fraction field is smoothed by a local vertex-averaging filter prior to curvature estimation, suppressing numerical instabilities associated with the large density jump at the interface.
We also account for the initial velocity jump that occurs in the drop at the first time-step for an incompressible solver.
As noted by \cite{marcotteDensityContrastMatters2019}, this is not a numerical artefact but a consequence of the conservation equations: the incompressible solver requires one time-step to establish the dipolar flow around the drop, during which the drop centroid acquires a finite velocity offset proportional to $\rhoo/\rhod$.
To suppress this transient, the drop-internal velocity is explicitly reset to zero after the first two time-steps, making sure the first two timesteps are sufficiently small to avoid losing any realistic deformation dynamics.

\subsection{Comparison of axisymmetric and 3D simulations with experiments}
\label{subsec:dropshape_exp_comparison}
The numerical setup used in this study, based on axisymmetric simulations in Basilisk, was extensively verified for spherical drops in \cite{parikThresholdDropFragmentation2025}.
In that work, the solver's performance was benchmarked against the canonical bag breakup experiment of \cite{Flock2012}, with results compared to both the experimental data and recent experiments and simulations in the literature.
One of the key findings of \cite{parikThresholdDropFragmentation2025} was that while axisymmetric simulations cannot capture late-stage dynamics and 3D phenomena such as bag rupture, they accurately model the initial deformation (at least up to the initiation time), correctly predicting the evolution of streamwise and transverse dimensions up to the formation of the critical pancake state ($t^* \approx 1.2$).
This reliability was confirmed across a broad parameter space of $\rho$, $\Oho$, and $\Ohd$ using 3D simulations (Appendix C of \cite{parikThresholdDropFragmentation2025}), predicting non-trivial pancake morphologies such as forward-pancake and the subsequent inflation orientation.
However, axisymmetric simulations were found to be unreliable for very low-viscosity systems ($\Oho,\Ohd \le 0.001$), where the ambient flow is chaotic and has low viscous dissipation in the drop fluid.
The parameters chosen for the present study lie well within the parameter space previously verified.

The purpose of this section is to extend this verification to the specific case of initially non-spherical (prolate and oblate) drops, which is the central focus of this work.
\begin{figure}
\centering
  \begin{subfigure}[c]{0.74\textwidth}
    \includegraphics[width=\linewidth]{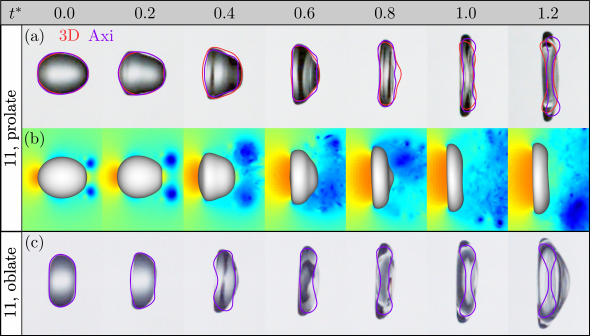}
    \label{fig:expsimcomp_vof-we11}
  \end{subfigure}
  \begin{subfigure}[c]{0.24\textwidth}
    \includegraphics[width=\linewidth]{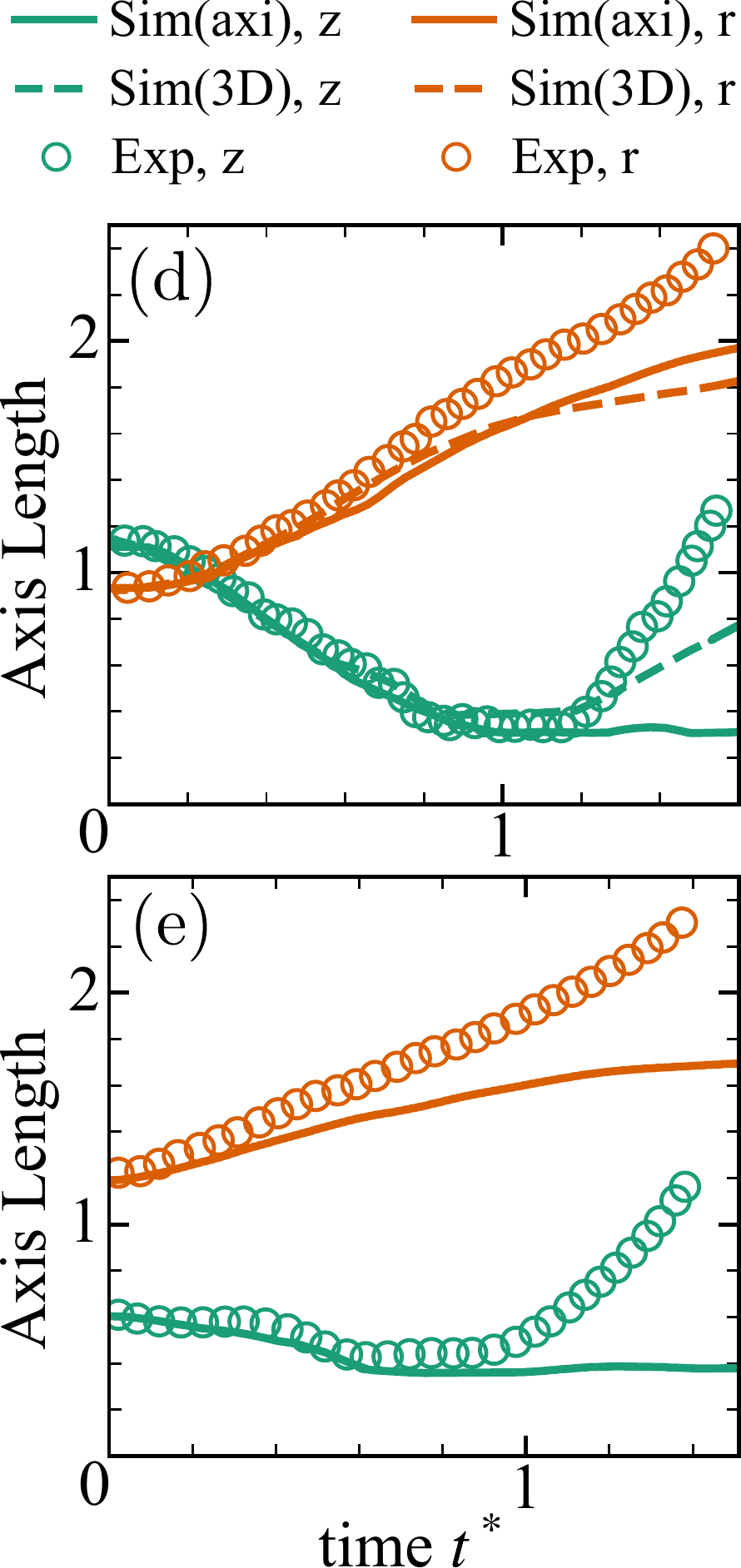}
    \phantomcaption
    \label{fig:expsimcomp_plots_we11}
  \end{subfigure}
  \vspace{-10pt}
  \caption
  {\label{fig:expsimcomp_we11}
    Comparison of experiments and simulations at $\We_0 \approx 11$ for prolate and oblate initial shapes.
    (a) Experimental shadowgraphs of the prolate drop, overlaid with the interface outlines from the 3D (red) and axisymmetric (purple) simulations, and (b) the corresponding 3D simulation renders.
    (c) Experimental shadowgraphs of the oblate drop, overlaid with the axisymmetric simulation outline; no 3D simulation was performed for the oblate case.
    (d,e) Time evolution of the streamwise ($z$) and transverse ($r$) axis lengths from the experiments and simulations for the (d) prolate and (e) oblate drops.
    Time ($t^*$) is scaled by the deformation timescale $\tau = \sqrt{\rho}(D/V_0)$.
  }
\end{figure}
To this end, we compare our simulations with a new set of experiments conducted using a water droplet (8 mm in diameter).
In this setup, the droplet falls freely under gravity along the vertical axis, which is also the axis of the upward-directed air jet.
Retracting the plate that obstructs the jet introduces a brief asymmetry in the loading until the plate is fully clear.
Once the plate is retracted, the drop is exposed to the full relative velocity $V_0$, and the subsequent aerodynamic loading is symmetric about the drop's axis of symmetry.
The droplets are released using a custom-designed mechanism \citep{fonnesbeckReleaseLargeWater2022, digheExtremelyLargeWater2024, almanashiLargeDropletRelease2026} that provides precise control over their initial shape at the moment of release, so that the subsequent free-fall oscillation is repeatable.
Details of the experimental setup, including the ramp time that establishes the impulsive exposure, are provided in appendix \ref{app:experimental_setup}, and a schematic of the setup is shown in Figure \ref{fig:expsetup}.
Following release, during free fall, the droplet undergoes natural oscillations.
Among the shapes traversed during these oscillations, the instants of maximum prolate elongation and maximum oblate flattening are selected as the starting states of impulsive acceleration, and the plate is retracted when the droplet reaches one of these two extreme shapes.
This timing ensures that both the oscillatory kinetic energy and the internal-flow contribution to interfacial shear are small at the onset of aerodynamic forcing, so that the initial condition is defined primarily by the drop shape.
The comparison simulations are therefore initialised with zero internal velocity, matching that experimental state, with the measured relative velocity imposed as $V_0$.
The dimensionless parameters for this system are a density ratio of $\rho = 815.9$, a drop Ohnesorge number of $\Ohd = 1.26\times 10^{-3}$, and an ambient Ohnesorge number of $\Oho = 6.96\times 10^{-4}$.

A wind velocity of approximately $6.6$ m/s results in an initial Weber number $\We_0 \approx 11$ for both initial shapes.
For the prolate case, the drop reaches its prolate extreme at a drop velocity of $2.37$ m/s, giving a relative velocity of $V_0 \approx 9.0$ m/s, with a streamwise axis length of $b = 9.32$ mm and a transverse axis length of $a = 7.64$ mm.
For the oblate case, the drop reaches its oblate extreme at a drop velocity of $2.03$ m/s, giving a relative velocity of $V_0 \approx 8.6$ m/s, with a streamwise axis length of $b = 5.10$ mm and a transverse axis length of $a = 9.71$ mm.
The rest initialisation is, however, less accurate for the oblate case than for the prolate one: while the drop centroid moves at approximately $2$ m/s, its rear face trails at approximately $1.7$ m/s, indicating residual internal velocities of order $0.3$ m/s that would contribute both oscillatory kinetic energy and interfacial shear, neither of which is represented in the comparison runs.
The initial shapes are realised a bit differently between the axisymmetric and the 3D simulations.
In the 3D simulations, they are realised as ellipsoids, with the streamwise and transverse axes set to $b$ and $a$, respectively, whereas the axisymmetric simulations are initialised with the true drop outline extracted from the experimental images.
This difference is incidental rather than deliberate.
The computationally expensive 3D simulations (approximately $96$ hours of wall-clock time across $128$ nodes on the Frontera supercomputer) were performed first, using the ellipsoidal idealisation, before a clean procedure for importing the true experimental outline into Basilisk was developed.
The inexpensive axisymmetric simulations were subsequently initialised with this true outline, while the already-completed 3D simulations retained the ellipsoidal shapes.
As the experimental outlines are only mildly non-ellipsoidal, as evident from the interface outlines in figure~\ref{fig:expsimcomp_we11}, this difference is expected to affect only finer details rather than the gross deformation outcome.
Both are initialised with zero internal velocities.
The experimental prolate and oblate extrema are close to, but not identical with, a finite-amplitude $(2,0)$ Rayleigh mode.
Each comparison run is therefore initialised from the experimental shape, represented either as an ellipsoid or as the extracted outline; the parametric study uses the volume-conserving Rayleigh modes of equation~\eqref{eq:rayleigh_shape_n0_general_main} (\cref{subsec:rayleigh_modes_corrected}).

Unlike the simulations that will be presented in the upcoming sections as part of the parameter sweep, gravity is retained in these comparison simulations to match the experimental conditions (see section~\ref{subsec:solver}).
A second set of experiments at a higher wind velocity (approximately $13$ m/s, $\We_0 \approx 32$) is presented in appendix~\ref{app:expsimcomp_we32}.

Figure \ref{fig:expsimcomp_we11} compares snapshots from these experiments with the corresponding renders from the current numerical simulations, along with a quantitative comparison of the drop's streamwise and transverse axis lengths.
For both the prolate and oblate initial shapes, an acceptable agreement between the experiments and the simulations is observed during the initial deformation phase, up to the formation of a pancake ($t^* \approx 0.8$).
The orientation of the pancake and the subsequent inflation are correctly predicted by the simulations, as can be seen from the similar positive slope of the streamwise lengths achieved by the simulations and the experiment.
For the prolate case, the axisymmetric and full 3D simulations remain in close agreement throughout this phase (panels a and d), justifying the use of axisymmetric simulations for the parametric study.
Both the true-shape axisymmetric and the ellipsoidal 3D simulations reproduce the flattening of the upstream pole and the rounder downstream pole that develop during the early deformation ($t^* \approx 0.2$).
At later times ($t^* \approx 0.6$), the axisymmetric simulation develops a spurious flattening of the downstream pole, an artefact of the axisymmetric constraint, which cannot resolve the asymmetry of the turbulent wake.
A similar observation was made by \cite{lingDetailedNumericalInvestigation2023}, who compared axisymmetric and 3D simulations directly at comparable Weber numbers and found that the two match very well up to $t^* \approx 0.5$ and remain in reasonable agreement up to $t^* \approx 0.8$.
They noted that the axisymmetric constraint produced an elevated pressure at the downstream pole that flattens it and pushes it upstream.
Overall, the axisymmetric simulation captures the deformation pathway from the initial shape to the pancake, including the asymmetry between the upstream and downstream halves of the pancake, and the rate of flattening $\dot{L}_z$ is reproduced well up to $t^* \approx 0.8$.

The oblate case, for which only an axisymmetric simulation was performed, shows comparable agreement (panels c and e); it reaches its minimum streamwise length at $t^* \approx 0.6$, much earlier than the prolate case ($t^* \approx 0.8$), and the simulations capture this trend.
A subtlety to keep in mind when comparing outlines is that every simulation outline, whether axisymmetric or 3D, is the trace of the interface in the meridional plane $\phi = 0$, with $\phi$ the azimuthal angle of equation~\eqref{eq:rayleigh_shape_nm}, whereas the experimental shadowgraph records the full projected silhouette of the drop.
For the prolate, convex shapes these largely coincide.
For the oblate drop, however, the two coincide only at the toroidal rim, so it is the rim geometry that provides the meaningful comparison; under this criterion the agreement is good.
At $t^* \approx 0.4$, the axisymmetric outline again shows the upstream-directed depression of the downstream pole, but the match recovers by the following snapshot at $t^* \approx 0.6$.
However, for both cases, the simulations underestimate the growth in transverse axis length during the pancake phase.
Beyond this point, as the drop begins to form a bag, the inherently three-dimensional dynamics cause the simulation results to diverge from the experimental data, as expected.

\begin{figure}
\centering
  \includegraphics[width=0.35\textwidth]{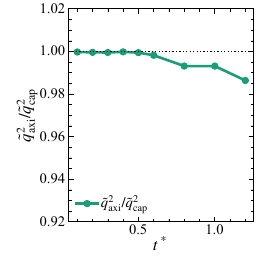}
  \caption{\label{fig:axi_frac}
    Axisymmetric fraction $\tilde{q}^2_\mathrm{axi}/\tilde{q}^2_\mathrm{cap}$: the share of total captured interface shape-variance residing in $m=0$ (axisymmetric) spherical-harmonic modes, extracted from the 3D simulation of the $\We\approx11$ prolate drop.
    This fraction remains above $98.6\%$ throughout the pre-bag regime ($t^*\le1.2$), confirming that nonlinear energy transfer to non-axisymmetric components is negligible and axisymmetric simulations are well-suited for tracking modal energy composition.
    Full details of the projection methodology are given in \cref{app:modal_analysis}.
  }
\end{figure}

This work aims to explore the role of initial shape of the drop on the subsequent deformation when exposed to impulsive acceleration.
For the purposes of this work, we require the numerical model to allow us to at least predict the essential deformation characteristics such as the shape and orientation of the pancake, since the drop's deformation and fragmentation characteristics are primarily determined by the balance of all the forces acting on the drop (and corresponding energetics) up until this point.
An accurate estimate of the final fragmentation morphology is not of interest for this work. Since the axisymmetric simulations capture the drop's evolution up to this critical stage, both for spherical \citep{parikThresholdDropFragmentation2025} and non-spherical initial shapes (verified here), the numerical model is deemed sufficient for the purposes of this investigation.
A concern with the use of axisymmetric simulations to study this problem could be that nonlinear interactions during deformation could transfer energy to non-axisymmetric mode components ($m \ge 1$), potentially invalidating the axisymmetric assumption. To address this, we verify this quantitatively, using the full 3D simulation of the $\We\approx11$ case described above. To do that, the axisymmetric fraction of the captured shape energy was quantified throughout the temporal evolution of the drop. The energy was found to remain above $98.6\%$ throughout the pre-bag regime ($t^* \le 1.2$); the key result is shown in figure~\ref{fig:axi_frac} and the full projection methodology used is detailed in \cref{app:modal_analysis}.

\section{Results}
\label{sec:results}

Consider a drop of volume-averaged diameter $D$ impulsively accelerated by a uniform flow of velocity $V_0$.
At any time $t$, the drop--ambient interface is denoted $\partial\Vol_d$, with $\bm{n}$ the outward unit normal pointing from the drop into the ambient medium.
The ambient-side stress tensor, comprising pressure and viscous contributions, is
\begin{equation}
  \label{eq:stress_tensor}
  \bm{T} = -p\bm{I} + \mu\twoD,
\end{equation}
where $\twoD = \bm{\nabla V} + (\bm{\nabla V})^T$ is the rate-of-deformation tensor.
The total hydrodynamic force exerted by the ambient medium on the drop is obtained by integrating this traction over the interface, and decomposes into pressure (form) and viscous contributions:
\begin{align}
  \label{eq:ambient_hydrodynamic_force}
  \bm{F}_{\mathrm{amb}}(t)
  & = \int_{\partial\Vol_d} \bm{T}\cdot\bm{n} \; dA,
  \\
  & = -\int_{\partial\Vol_d} p \bm{n} \; dA + \int_{\partial\Vol_d} \mu \twoD \cdot \bm{n} \; dA,
  \\
  & = \bm{F}_\mathrm{form}(t) + \bm{F}_\mathrm{visc}(t).
\end{align}
The instantaneous frontal area is $A_f = \pi(L_r/2)^2$, where $L_r$ is the transverse axis length, and the slip velocity is $U_\mathrm{rel} = V_0 - \Vcm$, where $\Vcm$ is the centre-of-mass velocity.
Using these, the form-drag coefficient is defined as
\begin{equation}
  \label{eq:form_drag_coefficient}
  C_{D,\mathrm{form}} = \frac{F_{\mathrm{form},x}}{\tfrac{1}{2} U_\mathrm{rel}^2 A_f},
\end{equation}
where the ambient density $\rho_o = 1$ in our non-dimensionalisation.
The viscous fraction $F_{\mathrm{visc},x}/F_{\mathrm{amb},x}$ measures the partitioning of drag between the two mechanisms.
As the Reynolds number increases and the flow develops, separates, and forms a wake, the drag progressively shifts from significant viscous-stress contributions to being predominantly controlled by the pressure differences associated with form drag.
Finally, we denote by $T_n = 2\pi/\omega_n$ the natural Rayleigh oscillation period of mode $n$, where $\omega_n$ is given by equation~\eqref{eq:Rayleigh_frequency}.

\begin{figure}
\centering
  \includegraphics[width=0.99\textwidth]{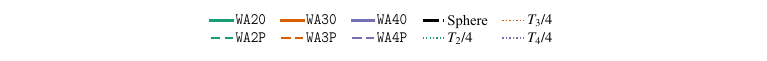}
  \par
  \begin{subfigure}[t]{0.32\textwidth}
    \insetfig{fig6a}{a}
    \phantomcaption
    \label{fig:waterair-Vcmz_t}
  \end{subfigure} \begin{subfigure}[t]{0.32\textwidth}
    \insetfig[83,88]{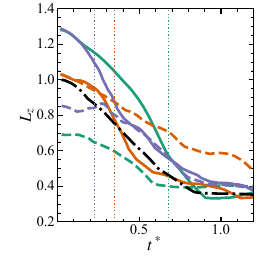}{b}
    \phantomcaption
    \label{fig:waterair-Lz_t}
  \end{subfigure}
  \begin{subfigure}[t]{0.32\textwidth}
    \insetfig{fig6c}{c}
    \phantomcaption
    \label{fig:waterair-Lr_t}
  \end{subfigure}
  \vspace{-10pt}
  \caption{\label{fig:waterair-kinetics}
  Temporal evolution of macroscopic kinematics and deformation metrics for a water drop impulsively accelerated in air.
  The panels display (a) the axial centre-of-mass velocity, $\Vcm$; (b) the drop's streamwise axis length, $L_z$; and (c) the transverse axis length, $L_r$, plotted against dimensionless time $t^*$.
  The results for the spherical reference case are compared against the three fundamental Rayleigh modes ($n=2, 3, 4$) initialised at their extreme phases ($0, \pi$).
  The vertical dotted lines mark the quarter Rayleigh period $T_n/4$ of each mode; panel~(a) carries an additional dashed line at the half-period $\Tthree/2 \approx 0.70$.
  }
\end{figure}

\begin{figure}
\centering
  \includegraphics[width=0.99\textwidth]{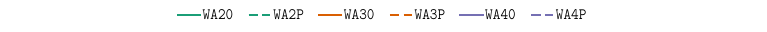}
  \par
  \begin{subfigure}[t]{0.32\textwidth}
    \insetfig{fig7a}{a}
    \phantomcaption
    \label{fig:waterair-Cdform_t}
  \end{subfigure} \begin{subfigure}[t]{0.32\textwidth}
    \insetfig{fig7b}{b}
    \phantomcaption
    \label{fig:waterair-viscfrac_t}
  \end{subfigure}
  \begin{subfigure}[t]{0.32\textwidth}
    \insetfig{fig7c}{c}
    \phantomcaption
    \label{fig:waterair-acm_t}
  \end{subfigure}
  \vspace{-10pt}
  \caption{\label{fig:waterair-aero}
  Temporal evolution of the axial aerodynamic loading for the same water drops as in figure \ref{fig:waterair-kinetics}, plotted against dimensionless time $t^*$.
  The panels display (a) the form-drag coefficient $C_{D,\mathrm{form}}$; (b) the viscous fraction of the axial ambient force, $F_{\mathrm{visc},x}/F_{\mathrm{amb},x}$; and (c) the centre-of-mass acceleration, $d\Vcm/dt^*$.
  As in figure \ref{fig:waterair-kinetics}, the three fundamental Rayleigh modes ($n=2,3,4$) are shown at their extreme phases ($0,\pi$), and the vertical dotted lines mark the quarter Rayleigh period $T_n/4$ of each mode.
  }
\end{figure}

We begin by examining the macroscopic kinematics for the six initial Rayleigh shapes relative to the reference spherical case, as presented in figure \ref{fig:waterair-kinetics}.
Figure \ref{fig:waterair-Vcmz_t} plots the temporal evolution of the centre-of-mass velocities, with the corresponding accelerations and axial drag breakdown shown in figure \ref{fig:waterair-aero}.
Initially, the $0$ phase of the $(2,0)$ mode (\cid{WA20}) exhibits the lowest centre-of-mass acceleration, achieving a $\Vcm$ approximately one-fifth the magnitude of the $\pi$ phase by $t^*\approx 0.4$.
This is consistent with its slender prolate shape and significantly smaller frontal area, and is reflected in its low form-drag coefficient together with the largest viscous fraction of all cases over $t^*\lesssim 0.3$ (figures \ref{fig:waterair-Cdform_t} and \ref{fig:waterair-viscfrac_t}).
From $t^*\approx 0.3$, its transverse length and hence frontal area begin to grow rapidly (figure \ref{fig:waterair-Lr_t}), and shortly after, at $t^*\approx 0.4$, \cid{WA20} undergoes a rapid increase in acceleration and surpasses \cid{WA2P} (figure \ref{fig:waterair-acm_t}).
This transition coincides with the development of a blunt upstream face, as the form-drag coefficient rises sharply while the viscous fraction collapses.
Since the form force scales with both the coefficient and the frontal area, the two grow together here and reinforce one another, so the jump in acceleration is steeper than the rise in $C_{D,\mathrm{form}}$ alone.

For the $(3,0)$ mode, the two phases share a similar projected frontal area.
At the very onset of acceleration ($t^*<0.05$), \cid{WA30} and \cid{WA3P} behave almost identically, with \cid{WA3P} marginally in the lead (figure \ref{fig:waterair-acm_t}).
Their paths separate immediately afterwards.
For \cid{WA30}, the viscous fraction collapses below $0.01$ by $t^*\approx 0.05$ (figure \ref{fig:waterair-viscfrac_t}), so form drag dominates from very early on, and its acceleration stays high and peaks near $t^*\approx 0.25$.
The streamlined \cid{WA3P} behaves quite differently.
It retains a larger viscous fraction of about $0.02$ over the same period, which points to a much smaller form drag.
Since its frontal area stays nearly constant until $t^*\approx 0.5$, the continuous fall of its form-drag coefficient over this interval (figure \ref{fig:waterair-Cdform_t}) corresponds directly to the flow becoming more streamlined around the drop, and its acceleration falls steadily in turn.
After $t^*\approx 0.5$, both the frontal area and the form-drag coefficient begin to rise, with the form-drag coefficient jumping precipitously to almost four times its earlier value, and the acceleration of \cid{WA3P} climbs rapidly to a roughly constant value of about $0.027$ (figure \ref{fig:waterair-acm_t}).
We return to the link between this behaviour and the modal oscillation timescale in \S\ref{subsec:discussions_waterair}.
A similar trend is observed for the $(4,0)$ mode.
The $0$ phase initially accelerates faster, consistent with its elevated viscous fraction during the early, low-form-drag stage (figures \ref{fig:waterair-viscfrac_t} and \ref{fig:waterair-Cdform_t}).
As the form-drag coefficient grows, its streamlined axial lobes give it the lower form drag of the pair relative to the $\pi$ phase, and consequently the lower acceleration at later times (figure \ref{fig:waterair-acm_t}).

Notable temporal modulations in centre-of-mass acceleration ($d\Vcm/dt$) are observed for the higher modes beyond $t^*\approx 0.4$.
For instance, \cid{WA30} starts with high acceleration but shows a distinct decrease and subsequent increase in $d\Vcm/dt$ at the quarter-period ($\Tthree/4 \approx 0.35$) and half-period ($\Tthree/2 \approx 0.70$, marked in figure~\ref{fig:waterair-Vcmz_t}), respectively.
Similar periodic variations corresponding to the Rayleigh time-period are observed for \cid{WA40}.
This indicates that centre-of-mass acceleration is governed not merely by the static aerodynamics of the instantaneous shape, but is dynamically modulated by the oscillation mode imposed on the drop.
The spherical case, lacking an imposed mode, shows no such fluctuations.

As the drops deform, the aerodynamic loading they experience evolves in complexity.
While the net aerodynamic force acts strictly in the streamwise direction (drag), the \textit{distribution} of pressure and shear stresses over the non-rigid interface performs work in both axial and radial directions.
Specifically, the strong pressure gradient from the upstream pole to the periphery generates a net radial forcing on the drop fluid.
Because the drop is deformable, this radial forcing performs work, effectively channelling a portion of the total aerodynamic power into radial expansion (internal oscillatory energy) rather than bulk acceleration (translational kinetic energy).
Thus, differences in initial shape do not merely alter the total drag coefficient; they fundamentally modulate how the external work is partitioned between accelerating the centre of mass and driving the radial deformation.
\cid{WA20} exhibits a significantly higher rate of radial expansion than all other cases, achieving a larger transverse length $L_r$ by $t^*\approx 0.8$.
Despite differing initial axial and radial lengths, most cases converge to a similar radial extent of $L_r \approx 1.45$ by $t^* \approx 1.0$, highlighting that different initial shapes undergo markedly different deformation rates to arrive at a similar pancake morphology.
The notable exception is \cid{WA3P}, which follows a distinct $L_z$ and $L_r$ evolution throughout the deformation: its radial length only begins to rise around $t^*\approx 0.5$, much later than the other cases, and consequently its $L_r$ remains the smallest and its $L_z$ the largest for the majority of the deformation process.

\begin{figure}
  \centering
  \includegraphics[width=0.7\textwidth]{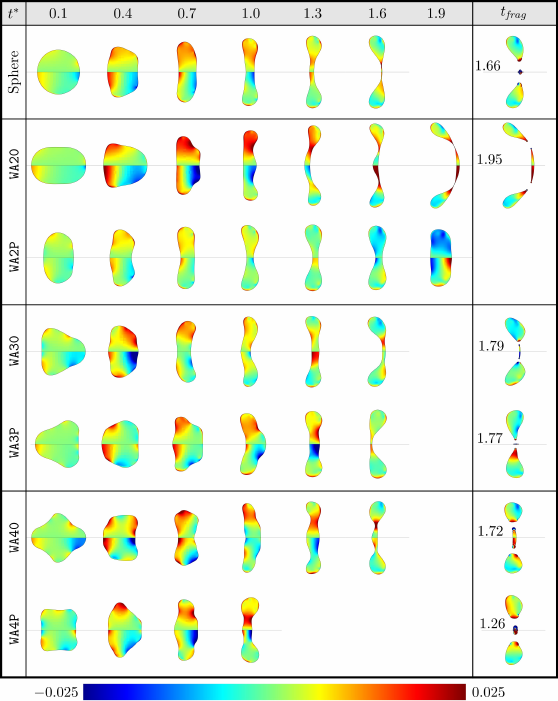}
  \caption{\label{fig:waterair-uin_comp}
  Temporal evolution of internal velocity fields at various $t^*$ for water drops initialised with different shapes.
  The upper half of each drop depicts the radial velocity component, while the lower half displays the axial velocity component.
  Snapshots beyond $t^* \approx 1.2$ (including the fragmentation column) are shown only to convey the \emph{relative} differences in deformation driven by the initial shape and the resulting modal coupling; they are not intended as physically realistic depictions of the real drop's late-time or fragmentation behaviour, which is inherently three-dimensional.
  }
\end{figure}

To elucidate how the macroscopic energy partitioning described above manifests within the fluid, we must isolate the internal motion from the drop's bulk translation.
Accordingly, we define an \textit{oscillatory velocity}, $\Vosc$, as the deviation of the local velocity field $\Vtot(\bm{x},t)$ from the volume-averaged centre-of-mass velocity $\Vcm$ (i.e., $\Vosc = \Vtot - \Vcm$).

Figure \ref{fig:waterair-uin_comp} presents the evolution of these internal velocities for the three initial shapes at various dimensionless times, $t^*$.
The snapshots utilise pseudo-colour renders partitioned into two halves: the upper half depicts the radial component ($\Voscr$), whilst the lower half displays the axial component ($\Voscz$).
The final column illustrates the drop's shape at the instant of fragmentation, where applicable.
With the sole exception of the $(2,0), \pi$ case, all initial shapes lead to fragmentation, albeit at varying times.
Since axisymmetric simulations cannot reproduce three-dimensional fragmentation morphologies or yield reliable absolute breakup times, the final-column snapshots and $t^*_\mathit{frag}$ values are retained solely to illustrate the \emph{relative} differences in deformation outcome across cases; because every case is subject to the same axisymmetric limitation, the relative ordering of outcomes remains meaningful.
The subsequent analysis focuses on the qualitative evolution of these internal flow patterns and their deviation from the spherical reference case.

For the $(2,0)$ mode, a striking difference in behaviour is observed between the two initial shapes.
Although \cid{WA20} is aerodynamically favourable---experiencing lower initial drag as noted in the macroscopic analysis---it exhibits significantly larger oscillatory velocities compared to the $\pi$ phase.
This apparent contradiction is resolved by considering the local pressure distribution rather than the net drag.
Analytical descriptions of impulsive deformation \citep{Villermaux2009,Kulkarni2014,parikThresholdDropFragmentation2025} indicate that radial expansion is driven by the pressure difference between the upstream pole and the periphery.
The prolate \cid{WA20} shape possesses a high curvature at the stagnation point (pole) and low curvature at the periphery, significantly enhancing this pressure gradient relative to the spherical or oblate cases.
This geometric advantage allows the $(2,0)$ mode to channel a larger fraction of the aerodynamic work into internal flow, driving strong deformation despite the lower global drag.
Conversely, \cid{WA2P} displays the lowest internal velocity magnitudes of all cases and notably fails to fragment.

However, the two extreme shapes of the $(3,0)$ mode exhibit very similar internal velocity structures, despite their differences in initial acceleration.
\cid{WA30} initially deforms faster, as visible in the snapshots at $t^* \approx 0.4$, driven by the steep pressure gradients associated with its blunt frontal profile.
A peak in radial velocity is observed where the prospective downstream peripheral bulge for the $\pi$ phase would reside, hinting at a strong influence of the modal oscillation on the internal oscillatory velocities.
Similarly, the $\pi$ extreme state (\cid{WA3P}) shows the highest radial velocities in the region where the upstream peripheral bulge is expected to form.
A key factor here is that the half-period of the $(3,0)$ mode is approximately $0.7$, which is very close to the time the reference spherical case takes to achieve its highest radial velocities.
The temporal evolution from $t^*=0.1$ to $1.0$ clearly shows both phases completing one half-cycle of their natural oscillation.
The drops fragment at $t^*_\mathit{frag} \approx 1.77$; while the exact fragmentation time cannot be trusted quantitatively given the limitations of the axisymmetric simulations beyond $t^*\approx 1$ (see \cref{subsec:dropshape_exp_comparison}), the half-period of the mode comfortably falls within the time to fragmentation even allowing for this uncertainty.
The two modes have thus undergone at least one full oscillation cycle by fragmentation; consequently, the drop experiences the full range of aerodynamic advantages and disadvantages associated with both shapes.

A more drastic difference in fragmentation time is observed for the $(4,0)$ mode drops, with the $\pi$ phase fragmenting significantly earlier than the $0$ phase.
The natural half-period of this mode is short (approximately $0.45$), implying the drop undergoes at least one full oscillation before fragmenting.
The initial flow for the $0$ phase, transitioning towards the $\pi$ phase, is constructive to the fragmentation process as it drives fluid away from the poles.
This is evident in the snapshots at $t^*=0.7$, where both the $0$ and $\pi$ phase drops display features resembling their opposite extreme shapes.
For \cid{WA40}, the highest radial and oscillatory axial velocities are observed where the mid-latitude bulge is expected to appear during the transition to the $\pi$ phase.
We observe similar behaviour in \cid{WA4P}, where the largest radial velocities are seen at the peripheral bulge at $t^* \approx 0.4$.
By $t^* \approx 0.8$, the two cases have theoretically achieved their opposite extremes and begin deforming back towards their initial shapes.
For \cid{WA4P} at $t^*\approx 1$, the drop has arrived at a physical morphology that supports fragmentation.
\cid{WA40}, on the other hand, is transitioning towards a shape with axial lobes at $t^* \approx 0.7$, and thus possesses kinetic energy that directly opposes bag formation.
Consequently, the drop requires an additional half-period to arrive at a similar quasi-equilibrium state with significantly less oscillatory kinetic energy opposing the formation of a bag.

\begin{figure}
  \centering
  \includegraphics[width=0.99\textwidth]{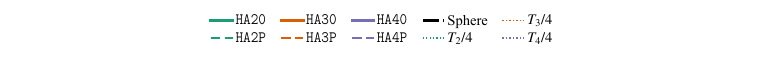}
  \par
  \begin{subfigure}[t]{0.32\textwidth}
    \insetfig{fig9a}{a}
    \phantomcaption
    \label{fig:retardantair-Vcmz_t}
  \end{subfigure} \begin{subfigure}[t]{0.32\textwidth}
    \insetfig[83,88]{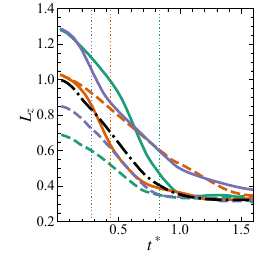}{b}
    \phantomcaption
    \label{fig:retardantair-Lz_t}
  \end{subfigure}
  \begin{subfigure}[t]{0.32\textwidth}
    \insetfig{fig9c}{c}
    \phantomcaption
    \label{fig:retardantair-Lr_t}
  \end{subfigure}
  \vspace{-10pt}
  \caption{\label{fig:retardantair-kinetics}
  Temporal evolution of macroscopic kinematics for a high-$\Ohd$ drop impulsively accelerated in air.
  The panels display (a) the axial centre-of-mass velocity, $\Vcm$; (b) the drop's axial length, $L_z$; and (c) the radial length, $L_r$, as functions of dimensionless time $t^*$.
  The spherical reference case is compared against the three fundamental Rayleigh modes ($n=2, 3, 4$) initialised at their extreme phases ($0, \pi$).
  }
\end{figure}

In contrast to the low-$\Ohd$ (water-air) case, a drop comprising fluid $100$ times more viscous exhibits markedly different behaviour under the same impulsive acceleration.
Since the ambient Reynolds number remains comparable to the water-air system ($Re \approx 0.8 Re_\mathrm{water}$), the external flow physics remains governed by the transition from initial viscous shear stress to flow separation and pressure drag.
We first examine the macroscopic kinematics in figure \ref{fig:retardantair-kinetics}.
Similar to the water-air case, \cid{HA20} initially possesses a more aerodynamic prolate shape with a smaller frontal area, resulting in a lower initial axial acceleration (figure \ref{fig:retardantair-Vcmz_t}).
However, as the drop deforms into a pancake and expands radially beyond the spherical reference, its frontal area increases and its shape becomes blunter; consequently, its axial acceleration grows, and its velocity eventually exceeds that of the reference case.

A more pronounced effect is observed for \cid{HA3P} and \cid{HA40}.
These cases also begin with shapes that are aerodynamically favourable relative to the sphere.
The lower aerodynamic forcing is accompanied by lower radial and axial deformation rates, as seen in figures \ref{fig:retardantair-Lz_t} and \ref{fig:retardantair-Lr_t}, where \cid{HA3P} and \cid{HA40} show almost no radial expansion until $t^* \approx 0.8$.
Beyond $t^* \approx 1$, these cases begin to exhibit comparable radial expansion rates as they finally approach a pancake morphology similar to the other cases.
However, the initial forcing disadvantage results in a significant "lag" in both their radial lengths and centre-of-mass velocities compared to the spherical reference.

Finally, we note that the mode-frequency associated modulations in acceleration, which were prominent in the water-air system, are significantly dampened here.
Figure \ref{fig:retardantair-Vcmz_t} shows smooth velocity profiles with minimal periodic fluctuations for the higher modes.
This absence of distinct acceleration peaks at $\Tthree/4$ or $\Tfour/4$ further confirms that the influence of the initial oscillation mode on the macroscopic kinematics is far less pronounced in the high-$\Ohd$ regime, where internal damping suppresses the dynamic coupling between the drop's vibration and its trajectory.

With the exception of the more aerodynamic cases (\cid{HA3P} and \cid{HA40}), most cases achieve very similar radial lengths of $L_r \approx 1.6$ by $t^* \approx 1.2$, as observed in figure \ref{fig:retardantair-Lr_t}.
Generally, cases with initially less aerodynamic shapes (excluding \cid{HA20}) exhibit perceptibly higher deformations and achieve thinner bags before retracting, consistent with the higher drag forces they experience during the early stages.
This contrasts with the water-air system, where constructive modal coupling drove unexpectedly large deformations even for unfavourable initial shapes.
The sole exception here is \cid{HA20}, which exhibits higher rates of deformation and achieves or exceeds the $L_r$ and $L_z$ values of the spherical reference.
\begin{figure}
    \centering
    \includegraphics[width=0.7\textwidth]{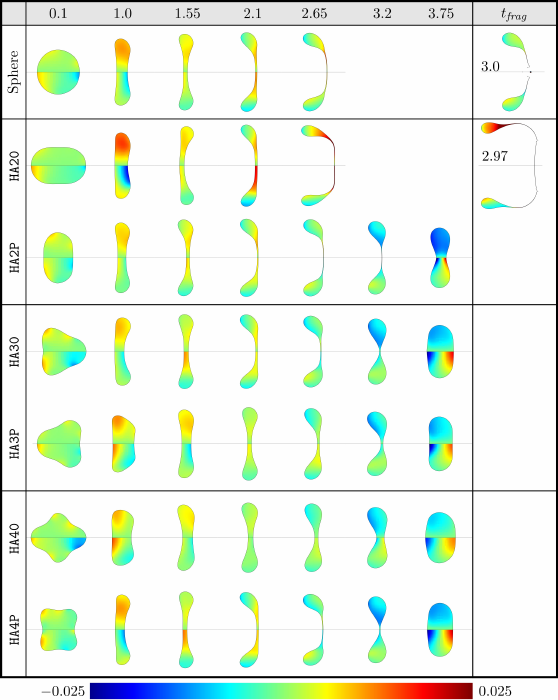}
    \caption{\label{fig:retardant-air-uin_comp}
    Temporal evolution of internal velocity fields for high-$\Ohd$ drops.
    The upper half of each snapshot depicts the radial velocity component, whilst the lower half displays the axial velocity component.
    The time sequence illustrates the enhanced viscous damping compared to the water-air system.
    As in figure~\ref{fig:waterair-uin_comp}, snapshots beyond $t^*\approx 1.2$ are included only to illustrate the relative shape-induced deformation differences, not as physically realistic depictions of the real drop's late-time or fragmentation behaviour, which is inherently three-dimensional.
    }
\end{figure}
The internal velocity fields, presented in figure \ref{fig:retardant-air-uin_comp}, reveal that the drops sustain much lower oscillatory velocities, $\Vosc$, compared to the water-air system -- an expected consequence of viscous damping smoothing internal velocity gradients.
Of all the cases simulated, only \cid{HA20} results in fragmentation, doing so at nearly the same time as the spherical reference case.
However, the resulting bag achieves higher inflation, and the internal oscillatory velocities are significantly larger relative to the sphere.

This unique outcome for \cid{HA20} underscores the critical importance of the pressure-driven mechanism identified in the water-air system.
While massive viscous dissipation suppresses the resonant modal coupling observed in the low-$\Ohd$ (water-air) regime, \cid{HA20} retains the distinct physical advantage of a large pressure difference between the high-curvature upstream pole and the low-curvature periphery.
This strong, geometrically induced pressure gradient is sufficient to drive radial expansion and overcome the viscous resistance, pushing the drop towards a shape topologically equivalent to \cid{HA2P}.
In contrast, \cid{HA2P}, which likewise starts from rest, lacks the requisite curvature-driven pressure gradient to initiate a strong internal flow.
Consequently, although both phases eventually arrive at a similar geometric stage, \cid{HA2P} lacks the kinetic energy required to progress to fragmentation.

For the higher modes, the role of the initial oscillation mode is largely diminished.
While $\Vosc$ peaks are still observed at regions corresponding to the subsequent geometric extremes (e.g., upstream mid-latitude regions for \cid{HA3P} and \cid{HA40}), the increased viscous dissipation effectively damps the feedback loop between internal flow and bulk deformation.
Modal resonance, which played a major role in the water-air system, is suppressed.
Instead, the initial aerodynamicity becomes the governing factor.
Initially less aerodynamic drops (e.g., \cid{HA30}, \cid{HA4P}) capture more aerodynamic work early on and consistently show greater deformation.
The $(2,0)$ mode remains the exception to this rule, where the non-aerodynamic $\pi$ phase unexpectedly fails to match the deformation magnitude of the spherical reference, purely due to the energetic deficit described above.

\begin{figure}
  \centering
  \includegraphics[width=0.99\textwidth]{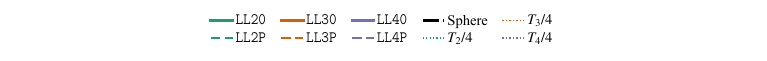}
  \par
  \begin{subfigure}[t]{0.32\textwidth}
    \insetfig{fig11a}{a}
    \phantomcaption
    \label{fig:liqliq-Vcmz_t}
  \end{subfigure} \begin{subfigure}[t]{0.32\textwidth}
    \insetfig[83,88]{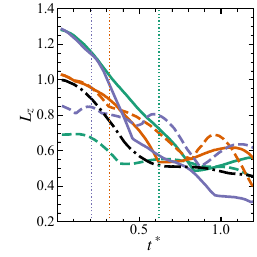}{b}
    \phantomcaption
    \label{fig:liqliq-Lz_t}
  \end{subfigure}
  \begin{subfigure}[t]{0.32\textwidth}
    \insetfig{fig11c}{c}
    \phantomcaption
    \label{fig:liqliq-Lr_t}
  \end{subfigure}
  \vspace{-10pt}
  \caption{\label{fig:liqliq-kinetics}
  Temporal evolution of macroscopic kinematics for a water-like drop in a liquid ambient ($\rho=10$).
  The panels display (a) the axial centre-of-mass velocity, $\Vcm$; (b) the drop's axial length, $L_z$; and (c) the radial length, $L_r$, as functions of dimensionless time $t^*$.
  Note the rapid acceleration of $\Vcm$ compared to the air systems.
  }
\end{figure}
\begin{figure}
    \centering
    \includegraphics[width=0.7\textwidth]{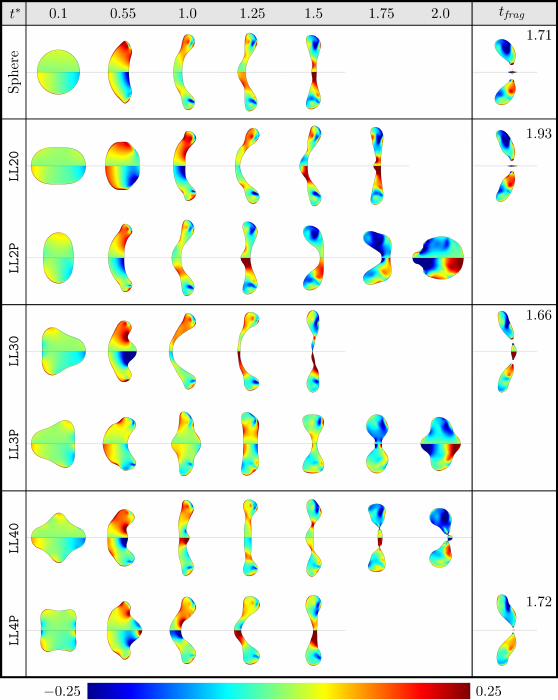}
    \caption{\label{fig:liqliq-uin_comp}
    Temporal evolution of internal velocity fields for a water-like drop in a liquid ambient.
    The upper half of each snapshot depicts the radial velocity component, whilst the lower half displays the axial velocity component.
    The scale of $\Vosc$ is significantly larger than in the air systems due to the lower inertia of the drop fluid.
    As in figure~\ref{fig:waterair-uin_comp}, snapshots beyond $t^*\approx 1.2$ are included only to illustrate the relative shape-induced deformation differences, not as physically realistic depictions of the real drop's late-time or fragmentation behaviour, which is inherently three-dimensional.
    }
\end{figure}

In the previous sections, the drop and ambient medium possessed very large viscosity and/or density contrasts.
Here, we consider the influence of the initial shape on a water-like drop impulsively accelerated in another liquid, where the density and viscosity are within an order of magnitude of the drop's properties ($\rho = 10.0, \Ohd = 1.258\times 10^{-3}, \Oho = 1.258\times 10^{-3}, \We = 10.0$).
The physics of this system is governed by two coupled effects that fundamentally alter the drag dynamics compared to the air systems.
First, the Ohnesorge number for the ambient phase is approximately double that of the water-air system ($Oh_{o,\mathrm{LL}} \approx 2 Oh_{o,\mathrm{WA}}$).
Since the Reynolds number scales inversely with the Ohnesorge number ($Re \propto Oh^{-1}$), the liquid-liquid system operates at an initial Reynolds number approximately half that of the air systems.
Second, due to the low density ratio, the drop accelerates rapidly, significantly reducing the instantaneous relative velocity ($U_\mathrm{rel} = V_\mathrm{freestream} - \Vcm$) and thus further depressing the instantaneous Reynolds number.

The combination of lower initial $Re$ and rapid acceleration (unsteady effects) delays the onset of flow separation.
Consequently, the flow remains in the \textit{attached viscous regime} for a much longer duration compared to the air systems.
In this attached-flow regime, pressure drag (form drag) is sub-dominant, and the total drag is governed primarily by viscous skin friction.
This creates an interesting inversion of the "aerodynamic" intuition established in the previous sections.
Streamlined shapes (such as \cid{LL3P} and \cid{LL40}), which typically minimise pressure drag, allow the flow to remain attached and accelerate over their curved surfaces, generating high integrated shear stresses.
In contrast, blunt shapes (like \cid{LL30}) possess larger stagnation zones with lower shear.
Thus, in this viscous-dominated early stage, the "more aerodynamic" shapes actually experience larger total drag forces.

This is quantitatively evident in figure \ref{fig:liqliq-Vcmz_t}.
Consistent with the viscous inversion hypothesis, \cid{LL3P} and \cid{LL40} exhibit slightly higher initial accelerations ($d\Vcm/dt$) than their blunt counterparts.
However, it is crucial to note that these differences are markedly smaller than those observed in the air systems.
With the exception of the $(2,0)$ mode, where a significant difference in frontal area between the prolate \cid{LL20} and oblate \cid{LL2P} still controls the drag forces, the velocity profiles for the higher modes remain tightly grouped during the start of the deformation process.
This prolonged similarity further supports the hypothesis that the surface areas exposed to viscous shear are comparable across these cases, and that flow separation which would drive a divergence in drag characteristics occurs much later.
This is the opposite of the water-air system, where the form-drag coefficients of the different phases separated strongly soon after release (figure \ref{fig:waterair-Cdform_t}) and drove the large early spread in their velocities.
As the flow eventually separates, the drag mechanism transitions back to being pressure-dominated.
This is observed for the $(3,0)$ mode, where \cid{LL3P} eventually falls behind \cid{LL30} in terms of centre-of-mass velocity, indicating that the blunt \cid{LL30} ultimately generates higher drag once a wake forms.
However, for the $(4,0)$ mode, \cid{LL40} remains at a higher centre-of-mass velocity relative to \cid{LL4P} throughout the simulation window, suggesting that the initial viscous advantage gained by the $0$ phase is preserved, or that its shape evolution delays the crossover to the pressure-drag regime.

Owing to the lower inertia of the drop fluid, approximately $10$ times larger oscillatory velocities are observed than in the $\rho=1000$ systems.
This is evident from the $\Vosc$ snapshots in figure \ref{fig:liqliq-uin_comp}, where radial ($\Voscr$) and axial ($\Voscz$) components as high as $0.25$ units are observed.

Notable differences arise compared to the water-air system.
While the fragmentation outcomes for the $(2,0)$ mode mirror the water-air system with only the $0$ phase (\cid{LL20}) achieving breakup, notable differences arise in the deformation magnitudes.
Specifically, the deformation of \cid{LL2P} is visibly lower compared to its water-air analogue \cid{WA2P}.
While \cid{LL20} reaches a radial length of $L_r \approx 1.5$ at $t^* \approx 1$, \cid{LL2P} achieves a similar radial length much earlier, at $t^*\approx 0.5$.
This difference implies that the effective duration of impulsive acceleration for \cid{LL2P} is reduced by approximately $0.5$ time units.
In such rapidly accelerating systems, this shortened interaction time results in a critical deficit in total energy input.
This explains the significant divergence in total deformation between the two phases, a contrast not observed in the water-air system.

This energy-deficit mechanism is further exemplified by the $(3,0)$ mode.
The more aerodynamic shape of \cid{LL3P} results in lower power input during the critical initial stages.
This difference in energy supply leads to a significant decrease in oscillatory velocities for \cid{LL3P} relative to \cid{LL30}, ultimately resulting in much lower deformation.
A similar observation is made for the $(4,0)$ mode, where \cid{LL4P}, being less aerodynamic than the spherical reference, is the only phase that captures sufficient energy to fragment.

\section{Discussion}
\label{sec:discussion}
Consider a drop of diameter $D$, density $\rhod$, viscosity $\mud$, and surface tension $\sigma$ subjected to impulsive acceleration.
At any time $t$, the internal velocity field is given by $\Vtot (\bm{x},t)$.
The volume-averaged velocity is denoted as $\Vcm$, representing the velocity of the centre of mass.
We define an \textit{oscillatory velocity}, $\Vosc = \Vtot - \Vcm$, which captures the internal motion of the fluid relative to the centre of mass.
Consequently, the oscillatory kinetic energy associated with the drop's deformation is given by
\begin{equation}
  \label{eq:oscillatory_kinetic_energy}
  \Kosc = \dfrac{1}{2} \rhod \int_{\Vol_\mathrm{d}} \Vosc \cdot \Vosc \; d\Vol = \Koscr + \Koscz,
\end{equation}
where $\Vol_\mathrm{d}$ is the drop volume.
The total kinetic energy $\Ktot$ decomposes into translational and oscillatory components (since the volume integral of $\Vcm \cdot \Vosc$ vanishes):
\begin{equation}
    \label{eq:Ktot2}
    \Ktot = \Kcm + \Kosc.
\end{equation}
The total instantaneous viscous power, $P_\mu$, and total energy lost to viscous dissipation at any time $t$, $E_\mu$, can be obtained as
\begin{equation}
  \label{eq:viscous_power}
  P_\mu = \dfrac{1}{2} \mu_d \int_{\Vol_d} \twoD:\twoD \; d\Vol
  \quad \& \quad
  E_\mu (t) = \int_0^t P_\mu \; dt.
\end{equation}
The surface potential energy is defined as the excess surface energy relative to the spherical equilibrium state, $\Epot = \sigma (A_d - A_{\mathrm{sph}})$, where $A_d$ is the instantaneous surface area and $A_{\mathrm{sph}} = \pi D^2$.
Thus, at time $t$, the total energy held in kinetic, potential, and heat energy by the drop is $\Etot = \Kosc + \Epot + \Kcm + E_\mu$, of which $\Eosc = \Kosc + \Epot$ is the recoverable oscillatory energy.

These metrics underpin the energetic analysis presented in this section.
The energy that drives them is supplied by the mechanical work performed on the drop by the surrounding ambient medium; we confirm that these diagnostics are physically meaningful by verifying the closure of the corresponding power and energy budgets in \cref{app:energy_power_closure}.

\subsection{Oscillatory kinetic energetics}
\label{subsec:discussions_waterair}
\begin{figure}
\centering
  \includegraphics[width=0.99\textwidth]{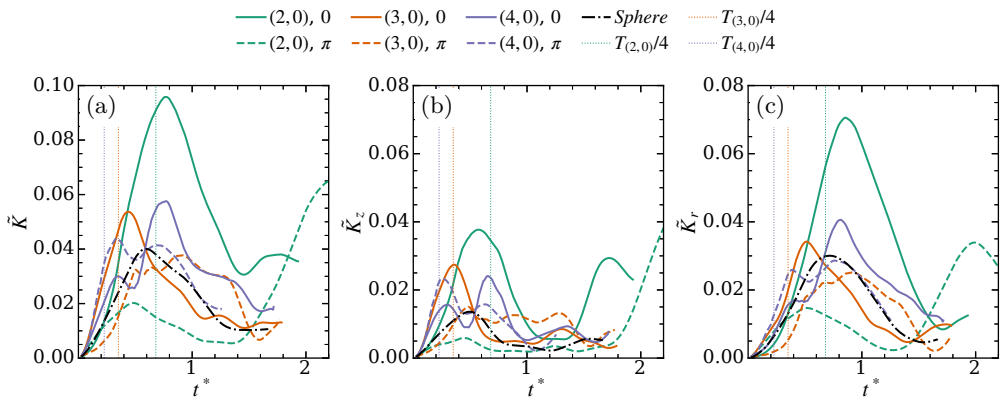}
  \par
  \begin{subfigure}[t]{0.32\textwidth}
    \insetfig{fig13a}{a}
    \phantomcaption
    \label{fig:waterair-KEosc_t}
  \end{subfigure} \begin{subfigure}[t]{0.32\textwidth}
    \insetfig{fig13b}{b}
    \phantomcaption
    \label{fig:waterair-KEoscz_t}
  \end{subfigure}
  \begin{subfigure}[t]{0.32\textwidth}
    \insetfig{fig13c}{c}
    \phantomcaption
    \label{fig:waterair-KEoscr_t}
  \end{subfigure}
  \vspace{-10pt}
  \caption{\label{fig:waterair-KEosc}
  Temporal evolution of (a) total oscillatory kinetic energy, (b) its axial component, and (c) its radial component for a water drop in air (\cid{WAxx}).
  }
\end{figure}

Figure \ref{fig:waterair-KEosc} presents the temporal evolution of the oscillatory kinetic energy $\Kosc$, its axial component $\Koscz$, and radial component $\Koscr$, for water drops in air (\texttt{WAxx}).
Each peak in figure \ref{fig:waterair-KEosc_t} represents an intermediate quasi-equilibrium state resulting from the superposition of the free modal oscillation and the forced deformation induced by the flow.
From the response of the spherical reference drop, we identify a characteristic timescale for the forced deformation, $\Tf \approx 0.6$, corresponding to the time required to reach the first peak in oscillatory kinetic energy.
This timescale $\Tf$ can be interpreted as the quarter-period of the forced oscillation driven by the impulsive acceleration.
Comparing $\Tf$ to the natural quarter-periods of the $(2,0)$, $(3,0)$, and $(4,0)$ modes reveals the quality of coupling between the forced and free deformation mechanisms.

The natural quarter-period of the $(2,0)$ mode is $\Ttwo/4 \approx 0.68$, which aligns closely with $\Tf$.
For \cid{WA20}, the transition from the $0$ phase (prolate) to the $\pi$ phase (oblate) requires internal flow from the poles to the periphery.
This is kinematically compatible with the flow induced by the impulsive acceleration, which flattens the drop into a pancake.
Consequently, \cid{WA20} experiences a \textit{constructive superposition}, where the modal and aerodynamic forces reinforce each other.
Driven by this physical compatibility and precise temporal alignment ($\Ttwo/4 \approx \Tf$), \cid{WA20} achieves the highest $\Kosc$ peak of all cases, effectively overcoming the disadvantage of its smaller, more aerodynamic frontal area.
Conversely, for \cid{WA2P}, the transition from $\pi$ to $0$ requires flow from the periphery to the poles, which directly opposes the aerodynamic forcing.
This \textit{destructive superposition} stifles the internal kinetic energy, resulting in a $\Kosc$ peak almost a factor of $2$ smaller than the spherical reference.
Despite the larger pressure forces experienced by the aerodynamically blunt \cid{WA2P}, the unfavourable modal coupling dominates, preventing fragmentation.

The $(3,0)$ mode introduces a timescale mismatch that fundamentally alters the coupling dynamics compared to the $(2,0)$ case.
With a natural half-period of $\Tthree/2 \approx 0.70$, the drop nearly completes a full transition between shape extremes within the forced oscillation timescale $\Tf \approx 0.6$.
For \cid{WA30}, the initial morphology presents a large frontal area to the free stream, generating significant pressure forces that drive fluid from the upstream pole towards the periphery.
This blunt upstream face produces a high and sustained form drag, as seen in figure \ref{fig:waterair-Cdform_t}, and the viscous fraction for \cid{WA30} collapses below $0.01$ within $t^*\approx 0.05$ (figure \ref{fig:waterair-viscfrac_t}).
The form-drag coefficient remains large until the drop begins relaxing towards its quasi-equilibrium shape near $\Tthree/4$.
Since the transition from the $0$ phase to the $\pi$ phase also involves fluid movement from the upstream surface to the downstream periphery, the two flow mechanisms initially complement each other.
However, the time to reach the quasi-equilibrium shape ($\Tthree/4$) is only about half the forcing timescale $\Tf$.
Consequently, as seen in figure \ref{fig:waterair-KEosc_t}, $\Kosc$ increases rapidly to reach a peak at $t^* \approx 0.50$---approximately the midpoint between $\Tthree/4$ and $\Tf$.
Because the two contributing oscillations do not synchronise their peak energies, the maximum $\Kosc$ achieved by \cid{WA30} is only modestly larger than that of the spherical reference, in stark contrast to the resonant amplification observed for \cid{WA20}.
Following this early peak, $\Kosc$ decays even before $t^* = \Tf$, mimicking the behaviour of the spherical case.

In contrast, \cid{WA3P} begins with a more streamlined upstream profile, resulting in lower pressure forces driving the peripheral expansion.
This streamlined profile keeps the form-drag coefficient low, and since its frontal area stays nearly constant until $t^*\approx 0.5$, the steady fall of $C_{D,\mathrm{form}}$ over this interval reflects a genuine streamlining of the flow around the drop (figure \ref{fig:waterair-Cdform_t}).
The viscous fraction also remains comparatively high, near $0.02$ (figure \ref{fig:waterair-viscfrac_t}).
Furthermore, the internal flow associated with the imposed surface oscillation initially opposes the aerodynamically driven deformation.
As a result, $\Kosc$ grows slower than the spherical case and maintains a lower value for most of $t^* < \Tf$.
However, due to its short modal oscillation period, the drop transitions to the $0$ configuration by $t^* \approx 0.7$, just past the forcing timescale $\Tf$.
At this stage, the drop exposes a larger frontal area to the flow while simultaneously possessing an internal velocity field that has become compatible with the external forcing.
The form-drag coefficient rises sharply in response, in step with the recovery of the acceleration of \cid{WA3P} (figure \ref{fig:waterair-aero}).
This physical configuration facilitates the transfer of external aerodynamic work into internal oscillatory kinetic energy, allowing $\Kosc$ to continue growing, albeit slowly, past $t^* \approx 0.7$.
A high $\Kosc$ is thus sustained over a longer period, resulting in a broader, plateau-like kinetic energy profile rather than the singular, high-amplitude peak observed for \cid{WA20} and \cid{WA30}.
This \textit{delayed constructive superposition} allows the $\pi$ phase to recover from its initial disadvantage and achieve sufficient deformation to trigger fragmentation.
The imposed mode is not erased by the violent deformation either.
As shown in the internal-velocity fields (figure \ref{fig:waterair-uin_comp}), the two phases carry lobes on opposite poles at $t^*\approx 1$, a clear signature of the initial $(3,0)$ oscillation surviving the deformation.

The $(4,0)$ mode possesses a much shorter quarter-period ($\Tfour/4 \approx 0.23$), allowing the drop to undergo a full transition to the opposite extreme and reach the next equilibrium state within the forcing timescale $\Tf$.
\cid{WA40} is initially more aerodynamic due to its axial lobes, reducing the pressure forces driving deformation.
While the internal flow from the $0$ to $\pi$ transition is directed from the poles to the periphery (constructive), the coupling is short-lived.
This is observed quantitatively in the first $\Kosc$ peak at $t^* \approx 0.4$ (between $\Tfour/4$ and $\Tf$).
The magnitude of this peak is relatively small because the modal oscillation begins its restorative phase early (at $\Tfour/4$), leading to destructive superposition within the first forcing period.
However, at $t^* \approx \Tfour/2$, the drop achieves a shape influenced by the $\pi$ state, which experiences significantly larger aerodynamic drag.
Consequently, when the drop reaches its next equilibrium state during the third quarter-period ($t^* \approx 3\Tfour/4$), the additional aerodynamic energy results in a much larger second $\Kosc$ peak as the oscillation accelerates back towards the $0$ phase.

Conversely, \cid{WA4P} begins with an aerodynamically unfavourable shape, experiencing large pressure forces at the onset of impulsive acceleration.
The forced oscillation contributes significantly to the drop's energy, resulting in a first $\Kosc$ peak that is significantly higher than that of \cid{WA40}.
Notably, this peak occurs at almost the same time for both cases.
Thus, \cid{WA4P} arrives at a similar pancake state with significantly larger kinetic energy, leading to earlier fragmentation.
In fact, the fragmentation times for \cid{WA40} and \cid{WA4P} differ by approximately $\Tfour/2$, suggesting that the delay in fragmentation corresponds directly to the half-period lag required for \cid{WA40} to achieve a comparable energetic state.
While absolute fragmentation times from axisymmetric simulations must be treated with caution, this comparative analysis highlights a strong physical connection between the initial oscillation mode and the resulting breakup dynamics.
\begin{figure}
\centering
  \includegraphics[width=0.99\textwidth]{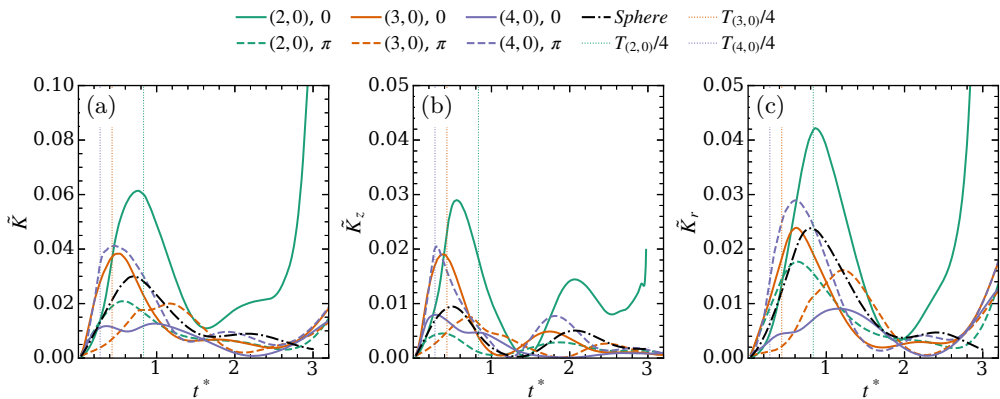}
  \par
  \begin{subfigure}[t]{0.32\textwidth}
    \insetfig{fig14a}{a}
    \phantomcaption
    \label{fig:retardantair-KEosc_t}
  \end{subfigure} \begin{subfigure}[t]{0.32\textwidth}
    \insetfig{fig14b}{b}
    \phantomcaption
    \label{fig:retardantair-KEoscz_t}
  \end{subfigure}
  \begin{subfigure}[t]{0.32\textwidth}
    \insetfig{fig14c}{c}
    \phantomcaption
    \label{fig:retardantair-KEoscr_t}
  \end{subfigure}
  \vspace{-10pt}
  \caption{\label{fig:retardantair-KEosc} Total oscillatory kinetic energy, and its axial and radial components for a high viscosity drop in air are presented in figures (a), (b), and (c), respectively. }
\end{figure}

Let us now consider the role of initial shape on the kinetic energetics of a high-viscosity drop in air as shown in the plots in figure \ref{fig:retardantair-KEosc}.
Most of our arguments and explanations follow the description provided for the case of a water drop in air.
The reference spherical case achieves its equilibrium state at $\Tf \approx 0.7$, when its oscillatory kinetic energy reaches a local maximum.
Apart from the reference spherical case, the only other case which fragments is \cid{HA20}, which undergoes a constructive superposition with the imposed impulsive acceleration.
\cid{HA20} achieves its peak $\Kosc$ at $t^* \approx 0.75$, which is unsurprisingly at approximately the mid-point of $\Tf$ and $\Ttwo/4 \approx 0.82$ for the high viscosity drop.
Crucially, however, the prolate initial shape generates a strong pressure gradient between the pole and periphery that persists throughout the deformation.
This geometric advantage allows \cid{HA20} to channel sufficient energy into radial expansion to overcome the massive viscous resistance and achieve fragmentation.

Since the drop fluid has approximately two orders of magnitude higher viscosity compared to a water drop, we expect to see much higher energy losses due to viscous dissipation and thus smaller $\Kosc$ peaks.
This can be inferred from figure \ref{fig:retardantair-KEosc_t}, where all cases show much smaller peaks compared to the water-air case for the same external aerodynamic forcing.
The cases that deform the least --- \cid{HA2P} due to its strong destructive superposition, and \cid{HA3P} and \cid{HA40} due to their aerodynamically favourable shapes --- all exhibit peak $\Kosc$ values significantly lower than the spherical reference case.
The higher loss of energies also results in a larger time (and thus higher work required to be done by aerodynamic forces) to fragmentation or to reach similar deformation levels for the higher viscosity drop.

It is also observed that the higher modes such as $(3,0)$ and $(4,0)$ only exhibit a single prominent $\Kosc$ peak, since the rapid energy dissipation all but dissipates the initial oscillation mode imposed on the drop.
Compared to the spherical case, any initial oscillation mode imposed on the drop apart from \cid{HA20} would increase viscous dissipation since all other initial states would either involve a strong destructive coupling right in the first half-period (in case of \cid{HA2P}), or go through an entire oscillation cycle exhibiting both constructive and destructive coupling phases (in case of \cid{HA3x} and \cid{HA4x}).
Interestingly, \cid{HA30} and \cid{HA4P} cases show initial $\Kosc$ peaks much larger than the spherical case, yet they fail to fragment.
Similar to the case of water drops, the two cases initially have aerodynamically unfavourable shapes that experience much larger pressure forces.
Much higher work is done by the external medium on the drop.
For \cid{HA30}, a fairly robust constructive superposition between the free and forced oscillation occurs since the flow is mostly aligned with a direction of movement away from upstream pole and towards the periphery.
The appearance of a downstream axial lobe is not compliant with the formation of a pancake, which reduces the coupling efficiency.
For \cid{HA4P}, the flow of drop fluid towards the formation of peripheral lobes couples constructively with the aerodynamic deformation, whereas the two axial lobes do not.
The higher aerodynamic work done coupled with not so great flow superposition results in peaks higher than the spherical case, but not as high as \cid{WA20}.
However, as the drop continues to deform past its first equilibrium state, the additional viscous dissipation incurred due to the presence of competing internal flows overcomes any initial aerodynamic advantage, $\Kosc$ rapidly drops, and the drop does not fragment.
Both \cid{HA3P} and \cid{HA40} start with an aerodynamic disadvantage and the first peak, similar to water drop, is relatively small.
While in the case of a water drop, both cases ultimately recovered their initial disadvantage, for the high viscosity drop, its higher viscous dissipation results in a small second peak and neither fragments.
In summary, the high-viscosity regime filters out the complex modal dynamics seen in water drops.
Fragmentation is determined by the $(2,0)$ mode's unique ability to combine constructive superposition with a geometrically sustained pressure gradient that drives simple radial expansion, minimising the viscous penalty associated with more complex shape evolutions.

\begin{figure}
\centering
  \includegraphics[width=0.99\textwidth]{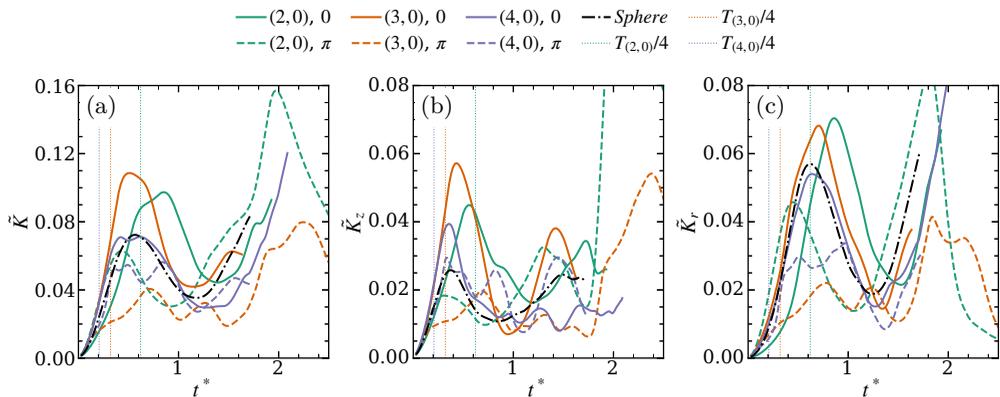}
  \par
  \begin{subfigure}[t]{0.32\textwidth}
    \insetfig{fig15a}{a}
    \phantomcaption
    \label{fig:liqliq-KEosc_t}
  \end{subfigure} \begin{subfigure}[t]{0.32\textwidth}
    \insetfig{fig15b}{b}
    \phantomcaption
    \label{fig:liqliq-KEoscz_t}
  \end{subfigure}
  \begin{subfigure}[t]{0.32\textwidth}
    \insetfig{fig15c}{c}
    \phantomcaption
    \label{fig:liqliq-KEoscr_t}
  \end{subfigure}
  \vspace{-10pt}
  \caption{\label{fig:liqliq-KEosc} Total oscillatory kinetic energy, and its axial and radial components for a water drop in another liquid are presented in figures (a), (b), and (c), respectively. }
\end{figure}

In the liquid-liquid system, the ambient density and velocity match the air cases, implying similar dynamic pressure scales.
However, the higher ambient viscosity ($Re \approx 0.5 Re_\mathrm{WA}$) maintains the flow in an attached, viscous-dominated regime.
Consequently, the drop experiences significant shear stresses on its upstream surface, which actively drive fluid towards the periphery.
This shear-driven mechanism sets up a faster internal flow compared to the pressure-dominated air systems, resulting in significantly higher oscillatory kinetic energies.
As shown in figure \ref{fig:liqliq-KEosc_t}, peak $\Kosc$ values are approximately $50\%$ higher than those in the water-air system.
Crucially, the low density ratio means the drop has very low inertia and accelerates rapidly towards the freestream velocity.
Since the aerodynamic forcing scales with the relative velocity squared ($\propto (V_{\mathrm{free}} - \Vcm)^2$), the available power drops precipitously with time.
This creates a narrow "energy window" during the initial stages.
Furthermore, it has been established that in such low-density systems where internal flow is driven by upstream shear, drops naturally deform into a forward-facing pancake (concavity pointing downstream) \citep{parikThresholdDropFragmentation2025}.
Thus, any initial oscillation that kinematically supports the formation of a forward-facing pancake will constructively superpose with the impulsive deformation.

The $(2,0)$ mode generally follows the trends seen in the air systems.
\cid{LL20} benefits from a highly efficient constructive superposition, where the transition from prolate to oblate aligns with the shear-driven flattening.
Conversely, while \cid{LL2P} (oblate) initially experiences larger total forcing due to its blunt profile, the internal flow required for its modal transition opposes the shear-driven deformation.
This destructive superposition prevents it from capturing sufficient energy before the drop accelerates and the external forcing wanes.
The efficiency of the constructive coupling for \cid{LL20} is remarkably quantitative; as seen in figure \ref{fig:liqliq-KEosc}, the difference in peak $\Kosc$ between \cid{LL20} and the spherical reference is almost exactly equal to the excess potential energy $\Epot(0)$ stored in the initial shape.

The $(3,0)$ mode offers a unique configuration for this shear-dominated regime.
For \cid{LL30}, the transition from the $0$ to $\pi$ phase involves moving fluid from the upstream surface to the downstream periphery.
This internal motion closely mimics the natural shear-driven flow field required to form a forward-facing pancake.
Consequently, \cid{LL30} experiences a resonance-like constructive superposition, resulting in the highest $\Kosc$ peak among all simulated cases---exceeding even \cid{LL20}.
In contrast, \cid{LL3P} begins with a streamlined shape that transitions in a way that is kinematically incompatible with the early shear forcing.
By the time the drop reaches its other extreme (blunt, $\pi$-like) and becomes receptive to aerodynamic work, it has already accelerated significantly.
With the relative velocity diminished, the drop cannot recover from this initial energy deficit and ultimately fails to fragment.

For \cid{LL40}, the free oscillation initially aids the shear-driven deformation, redistributing fluid from axial to mid-latitude lobes to support a forward-facing pancake.
This kinematic advantage leads to substantial early deformation; yet, the drop ultimately fails to fragment.
The short quarter-period of the $(4,0)$ mode ($\Tfour/4 \approx 0.2$) implies that the drop undergoes nearly a complete oscillation cycle within the characteristic timescale of the forced deformation.
Consequently, just as \cid{LL40} benefits early, it quickly transitions past this favourable quasi-equilibrium state.
As the drop evolves towards a phase characterised by a bulky central core, the internal flow actively opposes the thinning required for bag formation.
Once this window for efficient power input closes, the evolving configuration effectively curbs further energy capture.
In contrast, \cid{LL4P} successfully fragments despite starting with an incompatible kinematic state.
The rapid modal oscillation enables a quick transition into a suitable internal flow configuration by $t^* \approx \Tfour/2$; furthermore, the subsequent retraction phase supports the evacuation of fluid from the core, facilitating bag formation and ultimate fragmentation.

\subsection{Effect of initial mode on the fractional energy budget}
\label{subsec:energy_distribution}
\begin{figure}
\centering
  \includegraphics[width=0.99\textwidth]{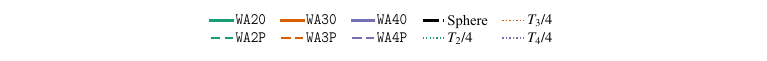}
  \par
  \begin{subfigure}[t]{0.32\textwidth}
    \insetfig{fig16a}{a}
    \phantomcaption
    \label{fig:waterair-Epot_KEOsc}
  \end{subfigure}
  \begin{subfigure}[t]{0.32\textwidth}
    \insetfig[83,88]{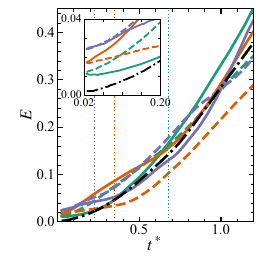}{b}
    \phantomcaption
    \label{fig:waterair-Etot_t}
  \end{subfigure}
  \begin{subfigure}[t]{0.32\textwidth}
    \insetfig{fig16c}{c}
    \phantomcaption
    \label{fig:waterair-KEoscbyEtot_t}
  \end{subfigure}
  \vspace{-10pt}
  \caption{\label{fig:waterair-energies}
  (a) Energy diagram plotting potential energy $\Epot$ versus oscillatory kinetic energy $\Kosc$.
  (b) Temporal evolution of total energy supplied to the drop $\Etot$.
  (c) Fraction of total energy converted to oscillatory kinetic energy $\Kosc/\Etot$.
  }
\end{figure}

Over the course of deformation towards fragmentation, the drops grow their surface areas by as much as a factor of $2$, thus superseding any direct contribution of the initial potential energy $\Epot(0)$ in the peak oscillatory kinetic energies gained by the drops.
This is evident from the energy diagram in the plot of potential energy against the oscillatory kinetic energy for a water drop in air as shown in figure \ref{fig:waterair-Epot_KEOsc}.
$\Epot$ grows by as much as an order of magnitude and $\Epot(0)$ is much smaller than the increase in peak $\Kosc$ relative to the spherical case, showing that the initially supplied additional surface energy $\Epot(0)$ cannot solely account for the additional $\Kosc$ in the $0$ phase.
From the energy diagram, we also note that except for the case that does not fragment, all other cases show almost constant or even a decrease in $\Epot$ and a simultaneous growth in $\Kosc$ during the deformation towards the quasi-equilibrium shape.
This hints at a preferential transfer of work done by the freestream towards amplifying oscillatory energy in the drop (instead of $\Epot$ for instance).
In fact, the primary mechanism through which an initial shape affects the energetics of the drop over the course of its deformation is by modulating the distribution of total work performed by the aerodynamic forces on the drop ($\Etot$).
The coupling between the initial mode and the impulsive acceleration determines how this external work is partitioned between translational kinetic energy ($\Kcm$), oscillatory energy ($\Kosc + \Epot$), and viscous dissipation ($\Emu$).

The total energy captured by the drop is shown in figure \ref{fig:waterair-Etot_t}.
During the initial stages ($t^*<0.5$), cases with less aerodynamic shapes, specifically \cid{WA2P}, \cid{WA30}, and \cid{WA4P}, capture energy most rapidly, as their large, blunt frontal areas allow pressure forces to perform significant work.
Conversely, \cid{WA3P}, with its streamlined initial profile, captures significantly less energy until $t^*\approx 0.4$, resulting in a net energy deficit of approximately $15\%$ compared to the other cases by $t^* \approx 1$.
However, for most cases, the differences in total energy supply ($\Etot$) are relatively small.
Thus, the drastic differences in fragmentation outcome are driven primarily by the efficiency of energy partitioning rather than the total magnitude of energy capture.
\begin{figure}
\centering
  \includegraphics[width=\textwidth]{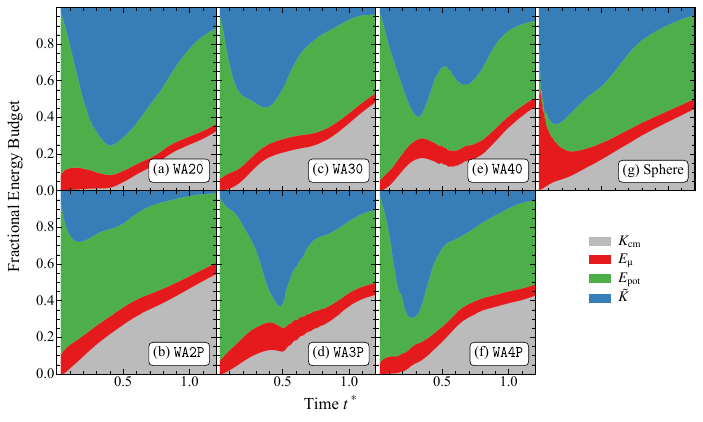}
  \caption{\label{fig:waterair-Epart_norm_t}
  Temporal evolution of the fractional energy budget for a water drop in air.
  The total energy is decomposed into translational kinetic energy ($\Kcm$), viscous dissipation ($\Emu$), additional surface energy ($\Epot$), and oscillatory kinetic energy ($\Kosc$), all normalised by the instantaneous total energy supplied to the drop $\Etot$.
  }
\end{figure}

Figure \ref{fig:waterair-Epart_norm_t} visualises this partitioning by plotting the various energy components as fractions of the instantaneous total energy $\Etot$.
Panel (a) clearly illustrates why \cid{WA20} achieves the highest oscillatory kinetic energies.
For $t^*<0.4$, \cid{WA20} channels almost all input energy into deformation ($\Kosc$) and viscous dissipation ($\Emu$), with negligible transfer to centre-of-mass acceleration.
This explains the "slow start" in $\Vcm$ observed in the Results section: the aerodynamic work is being utilised to deform the drop radially rather than accelerate it axially.
Consequently, \cid{WA20} holds nearly $80\%$ of its total energy in the oscillatory mode at its peak.
In contrast, \cid{WA2P} preferentially transfers energy to translational motion.
By $t^* \approx \Tf$, nearly $40\%$ of its total energy is locked in $\Kcm$, compared to less than $20\%$ for the other cases.
This leaves a significantly reduced fraction (only $\approx 20\%$) available for oscillatory kinetic energy.
Despite the larger total forces acting on the blunt \cid{WA2P} drop, this inefficient partitioning stifles the deformation.
Notably, the peak energy fraction for \cid{WA2P} occurs very early ($t^* \approx 0.15$), mimicking the spherical reference, whereas the resonant \cid{WA20} case sustains its energy growth until $t^* \ge 0.4$.

Finally, the role of viscous dissipation ($\Emu$), represented by the red region, reveals a non-conservative energetic cost of deformation.
While the cumulative dissipation is comparable across cases, the fractional rate of dissipation increases distinctly during periods of rapid $\Kosc$ growth (e.g., for \cid{WA40}).
This is expected, as establishing the strong internal velocity gradients required for oscillation inherently generates higher shear rates.
Thus, successful fragmentation requires a mode that not only captures aerodynamic energy but channels it into $\Kosc$ efficiently enough to overcome this viscous cost.
\begin{figure}
\centering
  \includegraphics[width=0.99\textwidth]{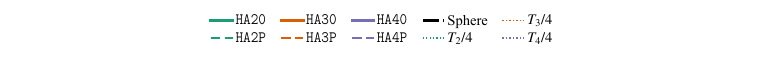}
  \par
  \begin{subfigure}[t]{0.32\textwidth}
    \insetfig{fig18a}{a}
    \phantomcaption
    \label{fig:retardantair-Epot_KEOsc}
  \end{subfigure}
  \begin{subfigure}[t]{0.32\textwidth}
    \insetfig[83,88]{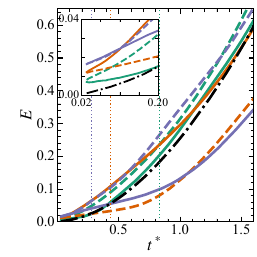}{b}
    \phantomcaption
    \label{fig:retardantair-Etot_t}
  \end{subfigure}
  \begin{subfigure}[t]{0.32\textwidth}
    \insetfig[83,88]{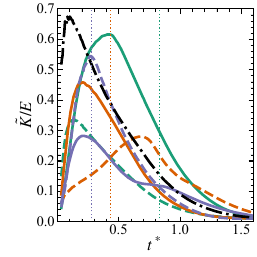}{c}
    \phantomcaption
    \label{fig:retardantair-KEoscbyEtot_t}
  \end{subfigure}
  \vspace{-10pt}
  \caption{\label{fig:retardantair-energies}
  (a) Energy diagram plotting potential energy $\Epot$ versus oscillatory kinetic energy $\Kosc$.
  (b) Temporal evolution of total energy supplied to the drop $\Etot$.
  (c) Fraction of total energy converted to oscillatory kinetic energy $\Kosc/\Etot$.
  }
\end{figure}
\begin{figure}
  \centering
  \includegraphics[width=\textwidth]{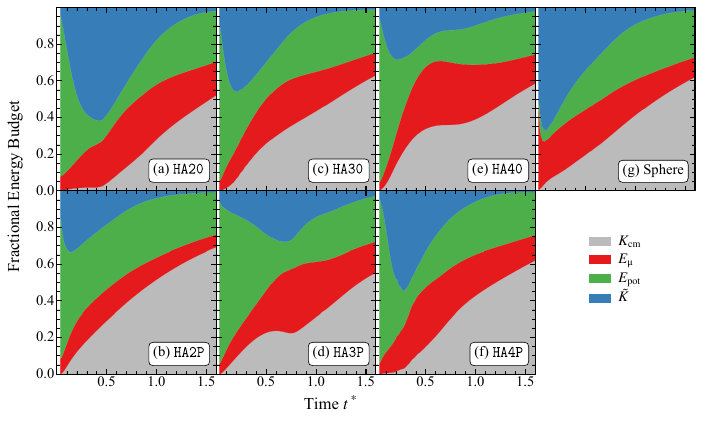}
  \caption{\label{fig:retardantair-Epart_norm_t}
  Temporal evolution of the fractional energy budget for a high-viscosity drop in air.
  The total energy is decomposed into translational kinetic energy ($\Kcm$), viscous dissipation ($\Emu$), additional surface energy ($\Epot$), and oscillatory kinetic energy ($\Kosc$), all normalised by the instantaneous total energy supplied to the drop $\Etot$.
  }
\end{figure}
For the high-viscosity drop, the energy supply shown in figure \ref{fig:retardantair-Etot_t} follows a similar trend to the water-air system, with all cases capturing comparable amounts of total energy during the initial deformation phase.
However, the partitioning of this energy is fundamentally altered by the viscous stress.
As evident in figure \ref{fig:retardantair-Epart_norm_t}, the viscous dissipation fraction ($\Emu$, red region) is substantial for all cases, representing a significant negative power flow on any internal motion.

Despite the significant energy losses, the fragmentation outcome ultimately is determined by how the remaining energy is allocated.
\cid{HA20} mimics the efficient partitioning of the spherical reference, channelling nearly $60\%$ of its total energy into the oscillatory mode ($\Kosc + \Epot$).
Crucially, it maintains a relatively low fraction of translational energy ($\Kcm$).
In contrast, \cid{HA2P} mirrors the behaviour of its low-viscosity counterpart (\cid{WA2P}), locking a large fraction of input energy into accelerating the centre of mass ($\Kcm$) rather than deforming the drop.
Due to the absence of strong internal oscillatory velocities, \cid{HA2P} actually incurs lower viscous dissipation than \cid{HA20}, yet it fails to fragment because the energy is essentially "wasted" on translation.

The cases \cid{HA30} and \cid{HA4P} present an interesting nuance.
Recall from figure \ref{fig:retardantair-Epot_KEOsc} that these cases achieve remarkably high absolute $\Kosc$ peaks, higher than the fragmenting sphere.
However, the fractional budget reveals why this energy is insufficient for breakup.
Neither case reaches the $60\%$ oscillatory fraction threshold seen in \cid{HA20}.
Furthermore, both cases exhibit a rapid expansion of the $\Emu$ region immediately following their peak $\Kosc$.
This indicates that the complex internal flows associated with these modes (e.g., competing lobed structures) generate severe velocity gradients that incur a higher viscous penalty than the simple radial expansion of \cid{HA20}.
Thus, high viscosity selectively dampens these complex modes, dissipating their energy before it can drive the drop to the fragmentation limit.

\begin{figure}
  \centering
  \includegraphics[width=0.99\textwidth]{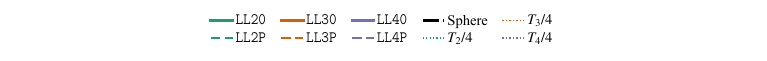}
  \par
  \begin{subfigure}[t]{0.32\textwidth}
    \insetfig[83,88]{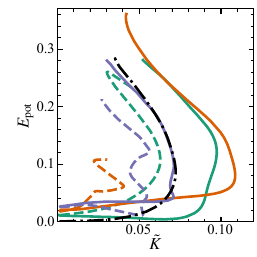}{a}
    \phantomcaption
    \label{fig:liqliq-Epot_KEOsc}
  \end{subfigure}
  \begin{subfigure}[t]{0.32\textwidth}
    \insetfig[83,88]{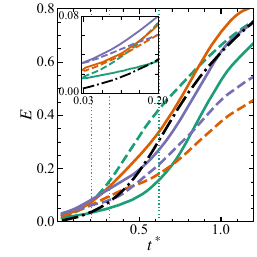}{b}
    \phantomcaption
    \label{fig:liqliq-Etot_t}
  \end{subfigure}
  \begin{subfigure}[t]{0.32\textwidth}
    \insetfig[83,88]{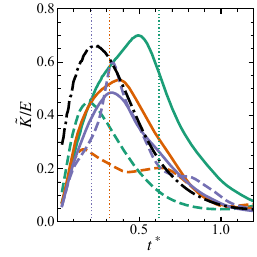}{c}
    \phantomcaption
    \label{fig:liqliq-KEoscbyEtot_t}
  \end{subfigure}
  \vspace{-10pt}
  \caption{\label{fig:liqliq-energies}
  Energy diagrams for a water drop impulsively accelerated in a similar liquid.
  }
\end{figure}

\begin{figure}
\centering
  \includegraphics[width=\textwidth]{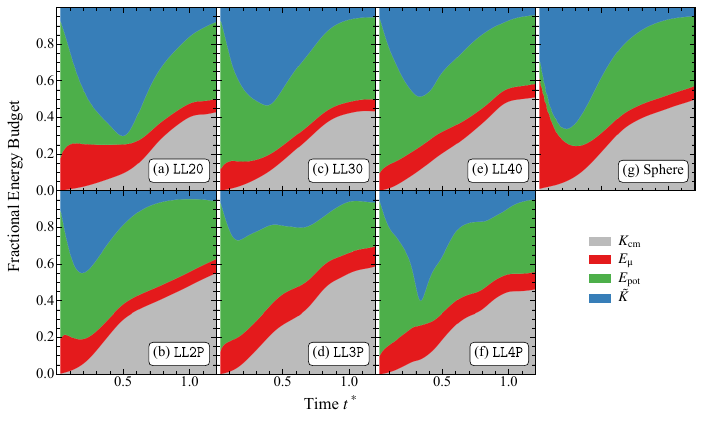}
  \caption{\label{fig:liqliq-Epart_norm_t}
  Temporal evolution of the fractional energy budget for a water drop in a liquid ambient.
  The total energy is decomposed into translational kinetic energy ($\Kcm$), viscous dissipation ($\Emu$), additional surface energy ($\Epot$), and oscillatory kinetic energy ($\Kosc$), all normalised by the instantaneous total energy supplied to the drop $\Etot$.
  }
\end{figure}

For a water drop impulsively accelerated in another liquid, the low density ratio strongly shapes the dynamics.
Because the drop has two orders of magnitude less inertia, it accelerates rapidly, reaching centre-of-mass velocities as high as $50\%$ of the freestream velocity, by the end of the deformation/fragmentation process.
Consequently, the aerodynamic power input diminishes dramatically at late times and hence the energy supplied by the ambient medium during the initial stages of the deformation becomes the dominant factor controlling the outcome.
In effect, the total energy captured by the oscillation during impulsive acceleration will primarily depend on the quality of coupling between the initial modes and the forced oscillation, and the amount of aerodynamic forcing acting on the drop, during these initial deformation stages.
This can be more clearly understood through the fractional energy budget plots for all the seven cases, including the reference spherical case, which informs us of the fragmentation threshold energy distribution.
This is shown in figure \ref{fig:liqliq-Epart_norm_t}.

The cases that fragment, i.e., \cid{LL20}, \cid{LL30}, and \cid{LL4P}, are distinguished by their ability to channel a larger fraction of early work into $\Kosc$.
All three maintain higher total oscillatory energies ($\Kosc+\Epot$) than the spherical reference throughout the deformation.
The budget for \cid{LL40} offers a final confirmation of the hypothesis proposed earlier.
During the first quarter-period, \cid{LL40} actually partitions a higher fraction of energy into $\Kosc$ than \cid{LL4P}, benefiting from its favourable initial coupling.
However, as it transitions past this optimal state, the coupling degrades.
This manifests in a smaller fraction of total energy being partitioned for $\Kosc$ at its peak, resulting in relatively less deformation for \cid{LL40}.
In contrast, \cid{LL4P} starts with a lower oscillatory fraction but rapidly improves its partitioning efficiency.
Because \cid{LL4P} achieves its peak efficiency while the relative velocity is still sufficiently high, it secures the net energy required for fragmentation, whereas the early advantage of \cid{LL40} is insufficient to sustain the deformation once the external forcing wanes.
However, we note that there are small differences in the coupling quality between the two phases and the amount of deformation is fairly similar.

\section{Conclusions and future research}
\label{sec:dropshape_conclusions}
In this study, we systematically investigate how the initial Rayleigh mode of a drop affects its deformation under impulsive acceleration.
The initial Rayleigh mode parametrises the initial shape and surface potential energy of a drop initialised at rest at the modal extremum.
For three different axisymmetric Rayleigh modes, VOF-based DNS simulations were used to analyse three distinct physical systems: a water drop in air, an enhanced viscosity water drop in air, and a water drop in another liquid of similar density. The key conclusions from this study are summarised below.

\begin{itemize}
    \item \textbf{Dynamic coupling dominates static aerodynamics:}
    The fate of a drop is not determined only by its initial drag coefficient.
    Instead, it is governed by the dynamic coupling between the drop's internal flow and the external aerodynamic forcing.
    When the internal flow induced by the initial mode (shape) aligns with the externally forced deformation, constructive superposition occurs and the drop captures more energy. This makes it more susceptible to fragmentation.
    Conversely, if the internal flow opposes the forcing, and destructive superposition occurs, the drop stabilises.

    \item \textbf{Resonance and timing in water-air systems:}
    A characteristic forcing timescale, $T_f \approx 0.6$, associated with impulsive aerodynamic deformation can be identified.
    The fundamental $(2,0)$ mode possesses a natural quarter-period ($\Ttwo/4 \approx 0.68$) that aligns closely with this forcing timescale.
    For the prolate ($0$) phase, this temporal synchronisation combines with a kinematically compatible internal pole-to-periphery flow to drive a resonant coupling that strongly promotes deformation toward fragmentation.
    For higher modes, the outcome depends on how their oscillation periods synchronise with $T_f$.
    A "recovery mechanism" was observed, where initially unfavourable shapes (like the $\pi$ phase of the $(4,0)$ mode) transition into a favourable configuration exactly when the aerodynamic forcing is strongest, whereas favourable shapes often transition out too quickly.

    \item \textbf{Viscous dissipation is paramount in high-viscosity drops:}
    For drops with relatively high-viscosity, internal viscous dissipation becomes the primary damping mechanism, suppressing the internal oscillatory flow and overriding the modal resonance effects seen in the water-air system.
    Higher modes ($n=3,4$) involve complex internal flow structures (e.g., competing lobes) that generate high shear rates and thus result in excessive viscous dissipation.
    This increased energy loss is only overcome for equally large aerodynamic and coupling advantages, the only such case being the \cid{HA20} case.
    The aerodynamic shapes experience lower drag, and they fundamentally lack the necessary aerodynamic power input to offset the increased viscous losses, and thus show visibly less deformation.

    \item \textbf{Shear-Driven deformation in liquid-liquid systems:}
    Liquid-liquid systems are characterised by low density ratios, and thus the drop accelerates rapidly, and the flow remains attached for a larger proportion of the deformation process.
    In these cases, shear stresses on the upstream face drive the deformation. This favours the formation of a ``forward-facing pancake". Initial modes that kinematically support this specific shape (like the $(3,0)$ mode), capture massive amounts of oscillatory energy. Due to acceleration of the drop at a rapid rate, the relative velocity decays quickly.
    This creates a small time-window of efficient power input.
    Drops that cannot capture energy early enough or those that transition out of a favourable state too quickly show less deformation.

    \item \textbf{Energy partitioning at the threshold:}
    Across all systems, the total energy supplied to the drop is often comparable.
    The determining factor for the deformation outcome is the efficiency of energy partitioning.
    Larger deformation, and ultimately fragmentation, occurs only when the drop channels a large fraction of the external work into oscillatory kinetic energy ($\Kosc$).
    Cases that show limited deformation typically ``waste" the energy by converting it into bulk translation ($\Kcm$) or losing it to viscous dissipation ($\Emu$).
\end{itemize}

In summary, the initial Rayleigh mode acts as a dynamic initial condition that sets the trajectory of the drop through the energy landscape.
Predicting secondary fragmentation requires accounting for this initial phase, as it can shift the outcome from catastrophic breakup to stability.

Two extensions naturally follow from the parameter space that has been explored in this study.
First, the oscillation amplitude was fixed at $\tilde A_n = 0.15$ for all simulations in this work.
Varying both the initial mode and its amplitude simultaneously would rapidly grow the parameter space, rendering a detailed exploration of modal coupling untenable.
Thus, the present study is deliberately reserved for comparing the modal coupling at a single, common amplitude.
The amplitude, however, directly sets the amount of energy that contributes to the coupling process, and is therefore expected to shift the fragmentation threshold to higher or lower $\We_0$, by an extent that is expected to scale with the amplitude.
Mapping this dependence is a natural next step, though very large amplitudes may not constitute viable initial states as the resulting shapes can be susceptible to a Rayleigh--Plateau (jet) instability under quiescent conditions.

Second, this study was restricted to near-threshold Weber numbers, where the initial mode is most influential.
As $\We_0$ is raised, the breakup morphology advances through a well-explored sequence --- from bag and bag-stamen, through multimode and shear-stripping (sheet-thinning), to the catastrophic regime \citep{raoSecondaryAtomizationDroplets2026}.
During bag and bag-stamen breakup, observed at $\We_0$ just above threshold, the aerodynamic forces impose a global deformation similar to the initial mode, so that the drop's response is governed by the modal coupling between the initial mode and the aerodynamic forcing, as explored in this work.
At higher $\We_0$, the breakup is instead driven by progressively smaller-scale interfacial instabilities, culminating in the Rayleigh--Taylor (RT) and Kelvin--Helmholtz (KH) instabilities that develop on the upstream face of the drop \citep{theofanousAerobreakupNewtonianViscoelastic2011}.
The growth of these instabilities is strongly controlled by the local curvature and velocity gradient at that face, which select the fastest-growing RT wavelengths near the upstream stagnation region and the KH wavelengths around the periphery \citep{theofanousAerobreakupNewtonianViscoelastic2011}.
The overall shape of the upstream face therefore dictates how this surface is partitioned between RT- and KH-susceptible regions: a flatter face presents a broad, low-curvature stagnation region that is prone to RT piercing, whereas a more streamlined face shrinks this region and shifts the balance toward KH-driven peeling around the periphery.
In fact, the formation of a flat upstream face is itself a necessary precursor to Rayleigh--Taylor piercing \citep{theofanousAerobreakupNewtonianViscoelastic2011}.
Crucially, the present results show that the initial mode sets the deformation pathway, and hence the upstream-face geometry that seeds this RT--KH partition.
We therefore expect its influence to persist well beyond threshold, throughout the shear-stripping regime, and to be erased only in the catastrophic regime, where near-instantaneous piercing of the entire liquid mass removes the imprint of any global mode.
Quantifying how a prescribed initial mode biases the RT and KH competition, and how long its influence survives the resulting multiscale cascade, would require fully three-dimensional simulations capable of resolving the azimuthal instabilities absent from the present axisymmetric framework.

\section{Acknowledgement} \label{sec:acknowledgement}
We wish to acknowledge the support of TACC and National Science Foundation for the compute time on Frontera under the Pathways program, which allowed us to conduct the computationally intensive simulations required for the paper.
We also thank Utah CHPC for the computation time on their clusters, which was also used to simulate some of the axisymmetric cases.

\vspace{0.5em}
Declaration of Interests: The authors report no conflict of interest.

\appendix

\section{Base-radius for volume-consistent finite-amplitude Rayleigh modes}
\label{app:derivation_b0}

\begin{figure}
    \centering
    \includegraphics[width=0.4\textwidth]{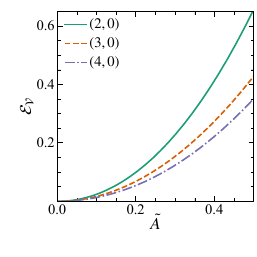}
    \caption{\label{fig:corrected_rayleigh_modes}
    Fractional volume excess $\mathcal{E}_V = (\Vol - \Vol_0)/\Vol_0$ introduced by the linearised Rayleigh shape equation~(\ref{eq:rayleigh_shape_n0_simple}) as a function of the $D$-normalised amplitude $\tilde{A} = A/(2R_0)$, for the $(2,0)$, $(3,0)$, and $(4,0)$ modes.
    At the working amplitude $\tilde{A}_n = 0.15$ the error reaches ${\approx}7\%$ for the $(2,0)$ mode; at $\tilde{A} = 0.30$ it reaches ${\approx}24\%$.
    }
\end{figure}

The commonly used linearised Rayleigh shape equation~(\ref{eq:rayleigh_shape_n0_simple}) is derived under the assumption of small (linearised) perturbations and implicitly identifies the base radius with the equilibrium sphere radius $R_0$.
Applied at finite amplitude, this parameterisation yields an initial drop shape whose enclosed volume exceeds the target sphere of radius $R_0$: the linearisation discards a quadratic correction that becomes significant once $\tilde{A}_n$ is not negligibly small.
The fractional volume excess $\mathcal{E}_V = (\Vol - \Vol_0)/\Vol_0$, where $\Vol_0 = \tfrac{4}{3}\pi R_0^3$, is plotted in figure~\ref{fig:corrected_rayleigh_modes} as a function of the $D$-normalised amplitude $\tilde{A} = A/(2R_0)$ for the three modes used in this study.
The error grows rapidly with amplitude, reaching values that would introduce a non-trivial mass discrepancy across the parametric cases; since both the Weber number and the drop inertia depend directly on the volume-averaged diameter, a consistent initial drop mass is essential for a meaningful parametric comparison.

This motivates retaining the base-radius degree of freedom $b_0 \neq R_0$ and solving for it from the exact volume constraint, as detailed below.
We aim to obtain a general form of $b_0$ for the superposition of any arbitrary number of Rayleigh modes and amplitudes.
We do this by equating the volume of the deformed drop for any and all superposed perturbations to the volume of a spherical drop of radius $R_\mathrm0$.
\begin{align}
  \label{eq:volume_conservation1}
  \dfrac{4}{3}\pi R_0^3 &= 2\pi \int_{\theta=0}^\pi \int_{r=0}^{r_n(\theta)} r_n^2\, dr\, \sin\theta\; d\theta,
  \\
                        &= \dfrac{2\pi}{3} \int_{\theta=0}^\pi \left[r_n(\theta)\right]^3\, \sin\theta\; d\theta,
\end{align}
\begin{subequations}
  \label{eq:volume_conservation2}
  \begin{alignat}{2}
  2R_0^3                &= \int_{\theta=0}^\pi \bigg[b_0 +&& \sum_{n=2}^{n\to\infty} A_n \; P_n (\cos\theta)\bigg]^3\, \sin\theta\; d\theta,
  \\
                        &= b_0^3 \int_{\theta=0}^\pi
                           \bigg[
                           1 
                         &&+ \dfrac{3}{b_0} \bigg( \sum_{n=2}^{n\to\infty} A_n \; P_n (\cos\theta) \bigg)
                         \\
                 & \quad &&+ \dfrac{3}{b_0^2} \bigg(\sum_{n=2}^{n\to\infty} A_n \; P_n (\cos\theta) \bigg)^2
                         \nonumber \\
                 & \quad &&+ \dfrac{1}{b_0^2} \bigg(\sum_{n=2}^{n\to\infty} A_n \; P_n (\cos\theta) \bigg)^3
                           \bigg]\, \sin\theta\; d\theta, \nonumber
  \end{alignat}
\end{subequations}
\begin{subequations}
  \label{eq:volume_conservation3}
  \begin{alignat}{2}
  2R_0^3                  &= b_0^3 \int_{\theta=0}^\pi
                           \bigg[
                           \sin\theta\; d\theta
                         &&+ \dfrac{3}{b_0} \bigg( \sum_{n=2}^{n\to\infty} A_n \; P_n (\cos\theta) \bigg) \sin\theta\; d\theta
                         \\
                 & \quad &&+ \dfrac{3}{b_0^2} \bigg(\sum_{n=2}^{n\to\infty} A_n \; P_n (\cos\theta) \bigg)^2 \sin\theta\; d\theta
                         \nonumber \\
                 & \quad &&+ \dfrac{1}{b_0^2} \bigg(\sum_{n=2}^{n\to\infty} A_n \; P_n (\cos\theta) \bigg)^3 \sin\theta\; d\theta
                           \bigg]. \nonumber
  \end{alignat}
\end{subequations}
The Legendre polynomials are orthogonal, which implies that the integrals of the product of two different Legendre polynomials over the interval $[-1,1]$ are zero, given that the polynomials are of different orders.
\begin{equation}
  \label{eq:orthogonality_legendre1}
  \int_{-1}^{1} P_n(x) P_m(x)\; dx = 
  \int_{0}^{\pi} P_n(\cos\theta) P_m(\cos\theta)\; \sin\theta\;d\theta = 0 \quad \text{for} \quad n \neq m.
\end{equation}
Setting $m=0$ in equation \ref{eq:orthogonality_legendre1} ($P_0(x) = 1$), we get:
\begin{equation}
  \label{eq:orthogonality_legendre2}
  \int_{0}^{\pi} P_n(\cos\theta) \; \sin\theta\;d\theta = 0 \quad \text{for} \quad 2\le n < \infty.
\end{equation}

We can also evaluate each of the integrals in equation \ref{eq:volume_conservation3} as follows:
\begin{equation}
  \label{eq:volume_conservationterm1}
 \int_0^\pi \sin\theta\; d\theta = 2,
\end{equation}
\begin{equation}
  \label{eq:volume_conservationterm3}
  \dfrac{3}{b_0^2} \bigg( \sum_{n=2}^{n\to\infty} A_n \; P_n (\cos\theta) \bigg)^2 \sin\theta\; d\theta = \dfrac{6}{b_0^2} \sum_{n=2}^{n\to\infty} \dfrac{A_n^2}{2n+1},
\end{equation}
\begin{equation}
  \label{eq:volume_conservationterm4}
  \dfrac{1}{b_0^3} \bigg( \sum_{n=2}^{n\to\infty} A_n \; P_n (\cos\theta) \bigg)^3 \sin\theta\; d\theta = \dfrac{2}{b_0^3} \sum_{i,j,k}^{\infty} A_i A_j A_k \begin{pmatrix}
               i & j & k \\
               0 & 0 & 0
              \end{pmatrix}^2.
\end{equation}
Here,
$\big( \begin{smallmatrix}
                i & j & k
             \\ 0 & 0 & 0
       \end{smallmatrix}$ \big)
is a specific case of the general Wigner 3-j symbol and evaluates to a scalar.
For $s=0.5(i+j+k)$, the Wigner 3-j symbol is given by:
\begin{equation}
  \label{eq:wigner3j}
  \begin{pmatrix}
               i & j & k \\
               0 & 0 & 0
              \end{pmatrix} = (-1)^s
                              \left[ \dfrac{s!}{(s-i)!\,(s-j)!\,(s-k)!} \right]
                              \sqrt{ \dfrac{ [2(s-i)]!\,[2(s-j)]!\,[2(s-k)]! }{ [2s+1]! } }.
\end{equation}
Substituting equations \ref{eq:volume_conservationterm1}, \ref{eq:volume_conservationterm3}, and \ref{eq:volume_conservationterm4} into equation \ref{eq:volume_conservation3} and simplifying gives us:
\begin{equation}
  \label{eq:volume_conservation5}
  2R_0^3 = b_0^3 \left[ 2
                      + \dfrac{6}{b_0^2} \sum_{n=2}^{n\to\infty} \dfrac{A_n^2}{2n+1}
                      + \dfrac{2}{b_0^3} \sum_{i,j,k}^{\infty} A_i A_j A_k \begin{pmatrix} i & j & k \\
                                      0 & 0 & 0
                      \end{pmatrix}^2
                 \right].
\end{equation}

We non-dimensionalise all length scales in equation \ref{eq:volume_conservation5} by drop diameter $D = 2R_0$.
Thus, $b_0 = \tilde b_0 D$ and $A_n = \tilde A_n D$ which results in:
\begin{equation}
  \label{eq:volume_conservation_final}
  1 = 8 \tilde b_0^3 
    + 24 \tilde b_0 \sum_{n=2}^{n\to\infty} \dfrac{\tilde A_n^2}{2n+1}
    + 8 \sum_{i,j,k}^{\infty} \tilde A_i \tilde A_j \tilde A_k
                              \begin{pmatrix} i & j & k \\
                                              0 & 0 & 0
                              \end{pmatrix}^2.
\end{equation}

Equation \ref{eq:volume_conservation_final} is a cubic equation in $\tilde b_0$ which can be solved using any standard numerical root-finding algorithm.
Since this is a one-time calculation, the computational cost of solving this equation is negligible.
We know that $\tilde b_0$ is bounded between $0$ and $0.5$ and hence the solution to equation \ref{eq:volume_conservation_final} is guaranteed using the bisection method.

It is instructive to compare equation \ref{eq:volume_conservation5} with the volume constraint derived by Rayleigh in Appendix~II of \citep{rayleighVICapillaryPhenomena1879}. Identifying the target sphere radius $R_0$ with Rayleigh's equilibrium radius $a$, the base radius $b_0$ with his $a_0$, and the amplitudes $A_n$ with his $a_n$, equation \ref{eq:volume_conservation5} can be rearranged as $R_0^3 = b_0^3 + 3 b_0 \sum_n (2n+1)^{-1} A_n^2 + \sum_{i,j,k} A_i A_j A_k \big(\begin{smallmatrix} i & j & k \\ 0 & 0 & 0 \end{smallmatrix}\big)^2$. The first two terms are precisely Rayleigh's equation~(34), obtained when the deviation from sphericity is small enough that $b_0 \approx R_0$ in the quadratic term and the cubic term is neglected. The final term is the finite-amplitude (cubic) contribution arising from $\int P_i P_j P_k \, \mathrm{d}\mu$, which the linearised analysis omits; retaining it makes the volume constraint exact at arbitrary amplitude.

\section{Experimental setup}
\label{app:experimental_setup}
\begin{figure}
\centering
\includegraphics[width=0.6\textwidth]{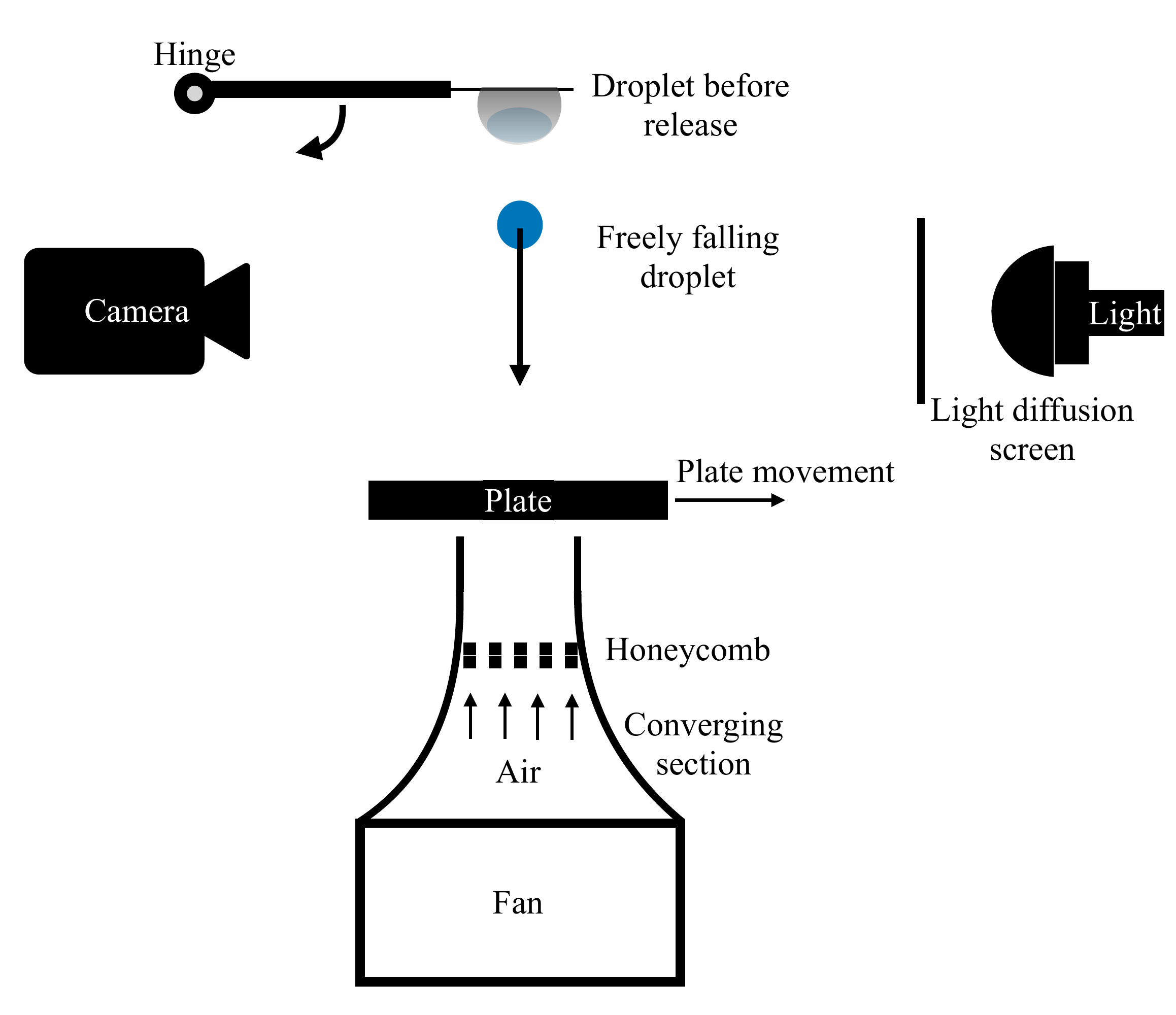}
\caption{\label{fig:expsetup}
Overview of the experimental setup designed to study the aerodynamic breakup of large droplets. The setup includes a custom droplet generation and release mechanism, a fan with a conical outlet and honeycomb flow straightener for air-jet formation, a sliding plate for impulsive exposure of the droplet to the jet, and the setup of the high-speed shadowgraphy system for visualisation.}
\vspace{\baselineskip}
\end{figure}

The experiments were conducted in the Splash Lab at King Abdullah University of Science and Technology (KAUST) using a custom-built air jet facility designed to investigate the aerodynamic breakup of large liquid droplets.
A schematic representation of the experimental setup is shown in Figure \ref{fig:expsetup}.
The setup comprises several key components: a novel droplet formation and release mechanism, an air jet system with a conical outlet, a movable plate assembly for impulsive exposure of the droplet to the air stream, and a solenoid-based triggering system.
The design and operation of the large droplet generation and release mechanism are described in detail in previous studies \cite{fonnesbeckReleaseLargeWater2022,digheExtremelyLargeWater2024,almanashiLargeDropletRelease2026}. These publications clearly show that the prolate and oblate drop shapes produced experimentally are repeatable. 
In each experiment, a single droplet is formed and released such that it falls vertically under the influence of gravity, undergoing natural $(2,0)$ mode shape oscillations during free fall.
The air jet used to induce aerodynamic breakup is generated by an AC Infinity fan (Model AL-CLT8) equipped with a custom-designed conical outlet.
A honeycomb structure positioned inside the cone serves to straighten the flow and reduce turbulence intensity.
The air jet velocity is measured using a hot-wire anemometer (Dantec MiniCTA 54T42), ensuring accurate characterisation of the airflow.

During the free fall of the droplet, the air jet is initially obstructed by a square sliding plate mounted on the inner guides of an aluminium frame.
The plate is held in a locked position above the air jet by rubber bands, maintaining tension until release.
Once the droplet reaches the desired oscillation extreme, the plate is rapidly retracted by actuating a solenoid, thereby exposing the droplet impulsively to the air jet.
The plate speed is typically $7$--$10$ m/s.
Estimating the ramp as the time to uncover a distance of order the drop diameter gives $t_{\mathrm{ramp}}\approx D/v_{\mathrm{plate}}\approx 1$ ms.
The corresponding nondimensional ramp time is $t^*_{\mathrm{ramp}}=t_{\mathrm{ramp}}/\tau$, where $\tau=\sqrt{\rho}\,D/V_0$ is based on the relative velocity $V_0$.
For the $\We_0\approx 11$ experiments of \cref{subsec:dropshape_exp_comparison}, $t^*_{\mathrm{ramp}}\approx 0.04$ for both the prolate and oblate shapes: a few percent of the deformation timescale, and short enough for the exposure to be regarded as impulsive.
The subsequent breakup process is recorded using high-speed shadowgraphy with a Phantom TMX-6410 camera operating at 10,000 frames per second. This imaging technique provides high temporal and spatial resolution necessary to capture the transient dynamics of droplet deformation and breakup.

\section{Comparison with experiments at a higher Weber number}
\label{app:expsimcomp_we32}
\begin{figure}
\centering
  \begin{subfigure}[c]{0.74\textwidth}
    \includegraphics[width=\linewidth]{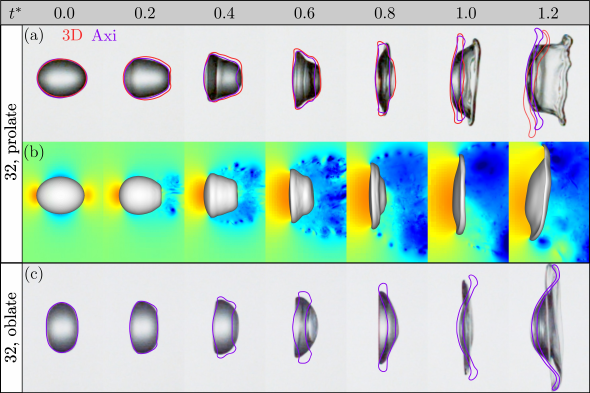}
    \label{fig:expsimcomp_vof-we32}
  \end{subfigure}
  \begin{subfigure}[c]{0.24\textwidth}
    \includegraphics[width=\linewidth]{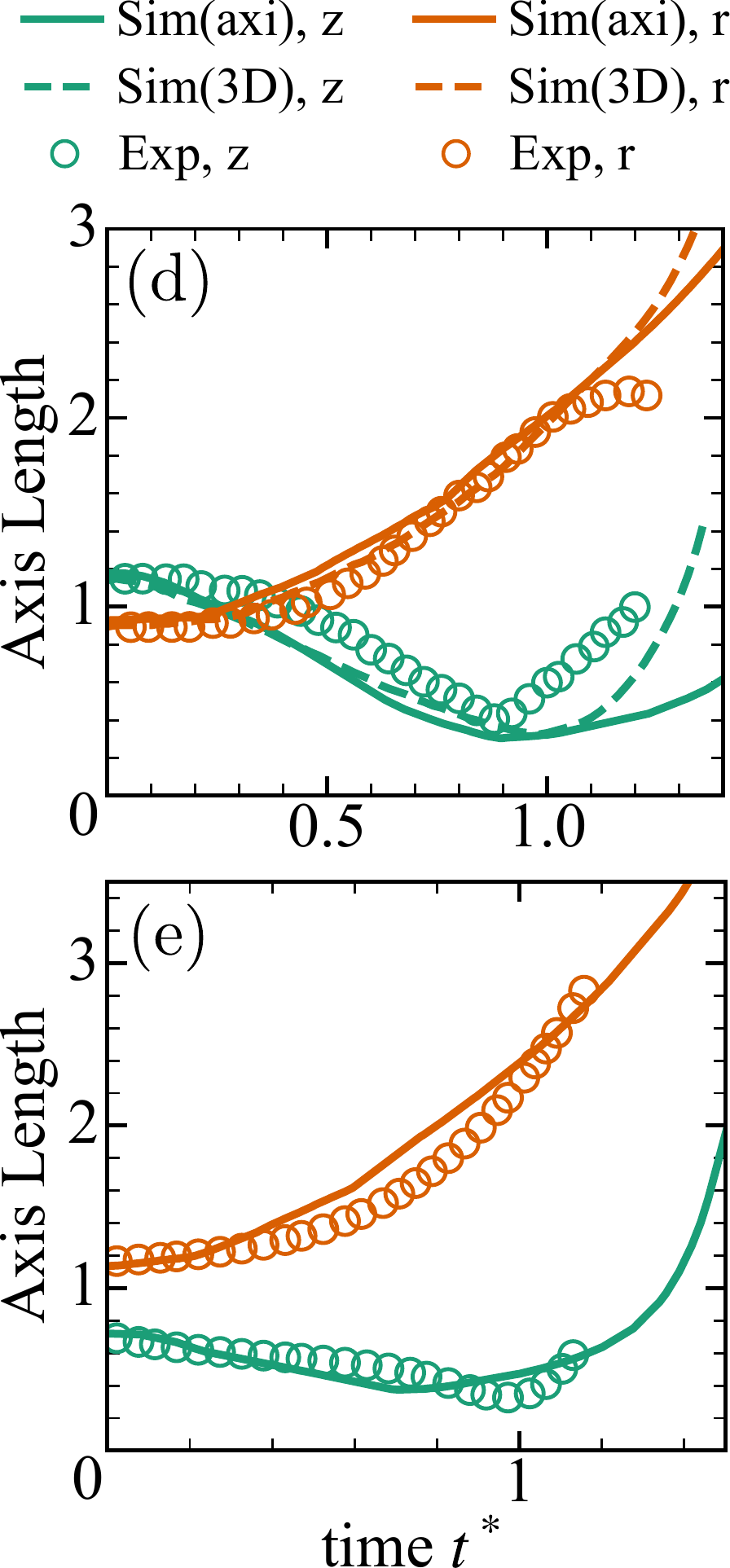}
    \phantomcaption
    \label{fig:expsimcomp_plots_we32}
  \end{subfigure}
  \vspace{-10pt}
    \caption{\label{fig:expsimcomp_we32}
    Comparison of experiments and simulations at $\We_0 \approx 32$ for prolate and oblate initial shapes.
    (a) Experimental shadowgraphs of the prolate drop, overlaid with the interface outlines from the 3D (red) and axisymmetric (purple) simulations, and (b) the corresponding 3D simulation renders.
    (c) Experimental shadowgraphs of the oblate drop, overlaid with the axisymmetric simulation outline (purple); no 3D simulation was performed for the oblate case.
    (d,e) Time evolution of the streamwise ($z$) and transverse ($r$) axis lengths from the experiments and simulations for the (d) prolate and (e) oblate drops.
    Time ($t^*$) is scaled by the deformation timescale $\tau = \sqrt{\rho}(D/V_0)$.
    }
\end{figure}

To complement the $\We_0 \approx 11$ validation presented in \cref{subsec:dropshape_exp_comparison}, we repeat the comparison at a higher wind velocity of approximately $13$ m/s for both the prolate and oblate initial shapes, corresponding to $\We_0 \approx 32$.
This Weber number lies well above the near-threshold regime that is the focus of the main text, where the initial drop shape is most consequential, so we present this comparison here rather than in \cref{subsec:dropshape_exp_comparison} to keep the main text focused on the primary Weber-number regime of interest.
For the prolate case, the drop reaches its prolate extreme at a drop velocity of $2.37$ m/s, giving a relative velocity of $V_0 \approx 15.3$ m/s, with a streamwise axis length of $b = 9.54$ mm and a transverse axis length of $a = 7.57$ mm.
For the oblate case, the drop reaches its oblate extreme at a drop velocity of $2.03$ m/s, giving a relative velocity of $V_0 \approx 15.1$ m/s, with a streamwise axis length of $b = 5.74$ mm and a transverse axis length of $a = 9.25$ mm.
The same physical ramp, $t_{\mathrm{ramp}}\approx 1$ ms (\cref{app:experimental_setup}), gives $t^*_{\mathrm{ramp}}\approx 0.07$ here, nearly twice the $\We_0\approx 11$ value, because $\tau$ shrinks with the larger $V_0$.
The higher-Weber-number experiments are therefore effectively less impulsive, and they deviate more from the simulations, which apply a perfect step jump in relative velocity at $t^*=0$.
As in the lower-Weber-number case, the 3D simulations use ellipsoidal initial shapes while the axisymmetric simulations are initialised with the true drop outline, and gravity is retained in both (\cref{subsec:dropshape_exp_comparison}).
Figure~\ref{fig:expsimcomp_we32} compares the experimental shadowgraphs with the simulation renders, along with the streamwise and transverse axis lengths.
At this higher Weber number the drop deformation becomes non-axisymmetric earlier than in the $\We_0 \approx 11$ case, so the axisymmetric idealisation breaks down sooner.
Even so, the overall behaviour is reproduced fairly well.
The initial pancake, including the asymmetry between its upstream and downstream halves, is captured, and the same axisymmetric artefact of the downstream pole being flattened and pushed upstream is again observed.
The subsequent shearing of the rim into a forward bag, with the accompanying sheet-thinning structures, is likewise reproduced by the axisymmetric simulations.
The bag dimensions, however, differ substantially from the experiments; we attribute this primarily to grid resolution and error tolerances that are insufficient to resolve the thin expanding sheet, rather than to the axisymmetric constraint itself.
Despite this, the streamwise and transverse axis lengths, $L_z$ and $L_r$, remain in reasonably good agreement throughout the deformation (panels d and e).
The various modelling caveats relevant to these comparisons are discussed in detail in \cref{subsec:dropshape_exp_comparison}.

\section{Convergence of higher order derived metrics}
\label{app:convergence}
\begin{figure}
\centering
  \includegraphics[width=0.99\textwidth]{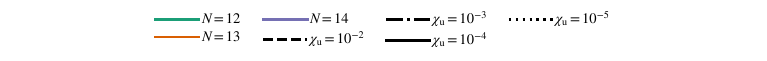}
  \par
  \begin{subfigure}[t]{0.32\textwidth}
    \insetfig[24,85]{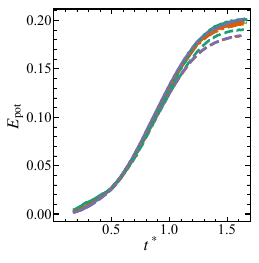}{a}
    \phantomcaption
    \label{fig:convergence-SE_t}
  \end{subfigure}
  \begin{subfigure}[t]{0.32\textwidth}
    \insetfig[24,85]{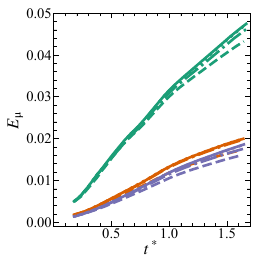}{b}
    \phantomcaption
    \label{fig:convergence-DE_t}
  \end{subfigure}
  \begin{subfigure}[t]{0.32\textwidth}
    \insetfig[26,85]{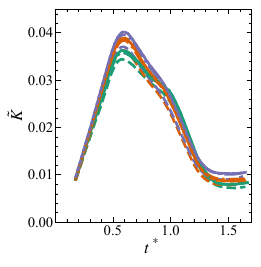}{c}
    \phantomcaption
    \label{fig:convergence-KEosc_t}
  \end{subfigure}
  \par
  \begin{subfigure}[t]{0.32\textwidth}
    \insetfig[26,85]{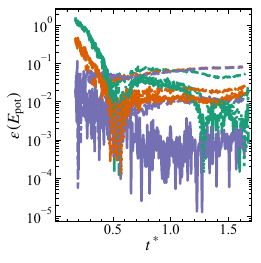}{d}
    \phantomcaption
    \label{fig:convergence-SE_err}
  \end{subfigure}
  \begin{subfigure}[t]{0.32\textwidth}
    \insetfig[26,85]{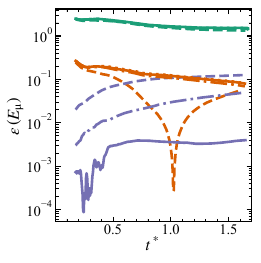}{e}
    \phantomcaption
    \label{fig:convergence-DE_err}
  \end{subfigure}
  \begin{subfigure}[t]{0.32\textwidth}
    \insetfig[26,85]{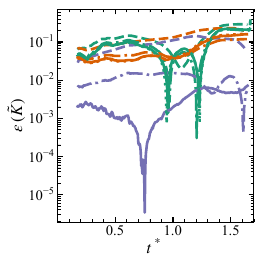}{f}
    \phantomcaption
    \label{fig:convergence-KEosc_err}
  \end{subfigure}
  \vspace{-10pt}
  \caption{\label{fig:convergence}
  Temporal development of additional Surface Energy ($\Epot$), cumulative viscous dissipation ($\Emu$), and oscillatory kinetic energy ($\Kosc$) are plotted for different values of maximum allowed refinement level $N$ (minimum allowed cell size), and wavelet error threshold for the velocity field $\chi_u$ are shown in panels (a), (b), and (c), respectively.
  Panels (d), (e), and (f) show the corresponding relative errors with respect to the best guess corresponding to $N=14$, $\chi_u=10^{-5}$.
  All three metrics converge for the parameters $N=14$, $\chi_u=10^{-4}$, as seen from their relative errors which remain below $10^{-2}$ for the entire duration.
  }
\end{figure}

As has been discussed in section \ref{sec:results} and \ref{sec:discussion}, higher order properties such as kinetic energy, surface energy, and viscous dissipation are metrics obtained from the simulations.
However, the convergence of first order statistics such as velocity field and volume fraction field does not ensure that second order statistics are also converged with respect to the grid resolution.
In order to test the convergence of these second order metrics, we simulate the impulsive acceleration of a water drop in air ($\rho\approx816$, $\Oho\approx0.0007$, $\Ohd=0.0013$, and $\We_0=12$) for $3$ different maximum refinement levels and $4$ different wavelet error thresholds for the velocity field.
Figure \ref{fig:convergence} illustrates all the convergence plots in the first row, i.e., panels (a), (b), and (c), and the corresponding relative errors with respect to the simulation with $N=14$, $\chi_u = 10^{-5}$ (i.e., the highest resolutions and strictest error thresholds tested).
For all metrics, the relative error remains below $10^{-2}$ for the parameters of choice for this work, i.e., $N=14$ and $\chi_u=10^{-4}$ confirming the convergence of energy and dissipation metrics obtained from the simulations.

\section{Power and energy budget}
\label{app:energy_power_closure}
The ambient traction forces (equations~\eqref{eq:stress_tensor} and~\eqref{eq:ambient_hydrodynamic_force}) form the source of mechanical energy input to the drop.
The instantaneous aerodynamic power acting on the drop, $P_\mathrm{amb}(t)$, and the total mechanical work performed by the ambient medium on the drop, $W_\mathrm{amb}(t)$, at time $t$, are given as
\begin{equation}
  \label{eq:ambient_power}
  P_{\mathrm{amb}}(t)
    = \int_{\partial\Vol_d} \left(\bm{T} \cdot \bm{n}\right)\cdot \bm{V} \; dA,
  \quad \& \quad
  W_\mathrm{amb}(t) = \int_0^t P_{\mathrm{amb}}(t') \; dt'.
\end{equation}
By application of Reynolds Transport Theorem on the volume enclosed by the drop fluid with respect to a static reference frame, we can derive the budget closure for energy and power transport through the control volume describing the drop.
The corresponding budget closures are
\begin{equation}
  \label{eq:energy_power_closures}
  \dot \Etot = P_\mathrm{amb},
  \quad \& \quad
  \Etot(t) - \Etot(0) = W_{\mathrm{amb}}(t).
\end{equation}
In this appendix, we verify that these energy and power budgets are sufficiently closed in the present Basilisk simulations.
This check is necessary because the energy diagnostics used in the main text rely on interfacial surface-area estimates, drop kinetic energy, and viscous dissipation calculations.

We define the normalised residuals for the power and energy budgets, $\ResP$ and $\ResE$, respectively, as
\begin{equation}
  \label{eq:energy_power_residuals}
  \ResP(t) = \dfrac{\dot\Etot(t) - P_\mathrm{amb}(t)}
                   {\left|\dot\Etot(\tau)\right|},
  \qquad
  \ResE(t) = \dfrac{\Etot(t) - \Etot(0) - W_\mathrm{amb}(t)}
                   {\left|\Etot(\tau) - \Etot(0)\right|}.
\end{equation}
Here, $\tau$ is the deformation timescale defined in \cref{eq:deformation_timescale}.
The values at $t=\tau$ are used only as fixed reference scales for non-dimensionalising the absolute residuals.
Thus, $\ResP$ measures the instantaneous mismatch between the measured rate of change of drop energy and the computed power input, while $\ResE$ measures the cumulative mismatch between the stored drop energy and the integrated power input.

For a perfectly closed budget, both residuals would be equal to zero.
However, machine-precision closure is not expected in Basilisk's VOF multiphase solver.
The geometrical VOF advection used in Basilisk is conservative for phase volume as well as momentum, where ``momentum conserving'' refers to conservation of the advective transport of the one-fluid mixture momentum \citep{pairettiBagModeBreakup2018}.
However, the discrete capillary force used by the continuum-surface-force formulation and the reconstructed PLIC surface area are not obtained from a single exactly energy-conserving discrete functional.
Thus, Basilisk does not ensure an explicit conservation of energy between the discrete capillary work and the measured change in reconstructed surface energy \citep{Popinet1999,Popinet2009}.
This produces a truncation-level residual that is expected to decrease under refinement, and is the dominant structural source of imperfect closure.

Additional residuals arise from the VOF representation of the interface.
The surface area entering $\sigma A_d$ is computed from piecewise-linear interface facets reconstructed from a compact stencil of the volume-fraction field using the Mixed-Youngs-Centred (MYC) method \citep{aulisaInterfaceReconstructionLeastsquares2007,Popinet2009}.
As the drop translates and deforms across the adaptive mesh, the local facet geometry changes discretely from cell to cell.
This produces small surface-area jitter.
The effect is usually weak in cumulative energy balances, but it is strongly amplified when $A_d(t)$ is numerically differentiated to compute $\dot\Etot$.
Thus a measure of power containing rate of change of surface energy would suffer from such high frequency jitter.

The ambient power terms introduce an additional post-processing error.
The pressure and viscous tractions are evaluated on the ambient side of the interface to avoid the interfacial mixed-phase region, where the one-fluid pressure contains the Laplace pressure jump \citep{Popinet1999,Popinet2009}.
The large density ratio and the filtering of the density and viscosity fields \citep{Popinet2009} further broaden the interfacial transition region by approximately one cell, which is why the ambient pressure is sampled with a finite stand-off from the reconstructed interface.
This one-sided reconstruction is necessary for obtaining a clean ambient traction, but it introduces a finite pressure stand-off, a least-squares reconstruction of the ambient velocity gradient, and a small mismatch between the sampling locations of pressure, velocity, and strain rate.
These approximations contribute directly to the residuals.

Smaller residuals also arise from finite grid resolution, AMR, and the approximate projection used by the centred Navier--Stokes solver.
Refinement and coarsening modify the discrete velocity and interface fields through prolongation and restriction.
In the approximate projection used by the centred solver, the projected face velocity used for fluxes is divergence-free by design, while the cell-centred velocity used in kinetic-energy diagnostics is only approximately divergence-free \citep{Popinet2009}.
Thus, the kinetic-energy diagnostics are not evaluated from the exact flux field used by the VOF and momentum advection steps.
This leaves a small bounded imprint of the divergence history in the pressure and velocity fields, which resets every timestep.

We also define a normalised residual for the axial centre-of-mass energy budget,
\begin{equation}
    \label{eq:cm_residual}
    \ResCM(t)
    =
    \dfrac{
    \Kcm(t) - \Kcm(0)
    -
    \int_0^t F_{\mathrm{amb},z}(t')\,\Vcm(t')\,dt'
    }
    {\left|\Kcm(\tau)-\Kcm(0)\right|}.
\end{equation}
This residual compares the ambient axial force with the corresponding change in centre-of-mass kinetic energy.
Unlike $\ResE$, it is largely decoupled from the deformation energetics, surface-energy reconstruction, and internal viscous dissipation.
It therefore provides an independent check of the force and velocity diagnostics used to compute the work done on the translating drop.

\begin{figure}
\centering
  \includegraphics[width=0.99\textwidth]{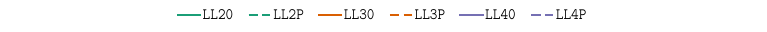}
  \par
  \begin{subfigure}[t]{0.325\textwidth}
    \insetfig[24,87]{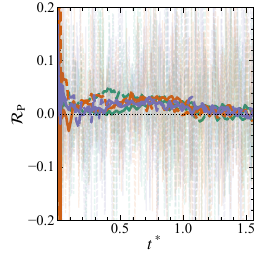}{a}
    \phantomcaption
    \label{fig:energypowerclosure-powererr_t}
  \end{subfigure}
  \begin{subfigure}[t]{0.325\textwidth}
    \insetfig[24,87]{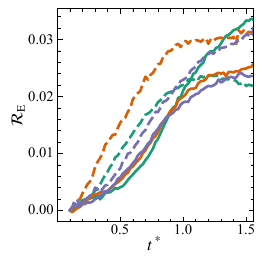}{b}
    \phantomcaption
    \label{fig:energypowerclosure-RE_t}
  \end{subfigure}
  \begin{subfigure}[t]{0.325\textwidth}
    \insetfig[24,87]{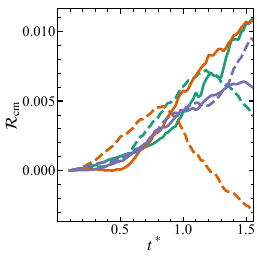}{c}
    \phantomcaption
    \label{fig:energypowerclosure-RCM_t}
  \end{subfigure}
  \vspace{-10pt}
  \caption{\label{fig:energy_power_closure}
  Closure of the instantaneous power budget, cumulative energy budget, and axial centre-of-mass energy budget for the water-drop-in-liquid-ambient case.
  (a) shows the normalised power residual $\ResP$.
  The raw signal contains high-frequency fluctuations caused primarily by differentiation of reconstructed surface-area jitter.
  The filtered signal shows the mean residual after applying a Savitzky--Golay filter.
  (b) shows the cumulative energy residual $\ResE$, and (c) shows the centre-of-mass residual $\ResCM$.
  The small cumulative residuals indicate that the energy transfer trends discussed in the main text are not caused by a systematic numerical energy imbalance.
  }
\end{figure}

In \cref{fig:energy_power_closure}, we plot these residuals for the system with a water drop in another liquid.
The instantaneous power residual in \cref{fig:energypowerclosure-powererr_t} exhibits high-frequency oscillations in the raw signal.
These oscillations are expected because $\ResP$ contains $\dot\Etot$, and direct differentiation amplifies the small PLIC facet-area jitter in $A(t)$.
After low-pass filtering using a Savitzky--Golay filter of polynomial order $2$, the residual fluctuates about zero without a secular drift.
This confirms that the power budget closes in the mean, and that the raw high-frequency fluctuations do not represent a persistent numerical source or sink of energy.

The cumulative residuals provide the stricter test of the net energy transfer.
As shown in \cref{fig:energypowerclosure-RE_t,fig:energypowerclosure-RCM_t}, both cumulative budgets remain small over the simulated interval.
The full drop-energy residual remains below approximately $3\%$ up to $t^*=1.5$, while the centre-of-mass residual remains near $1\%$.
The smaller value of $\ResCM$ is expected because this budget is dominated by axial translation and is less sensitive to surface-energy reconstruction, capillary work, and deformation-scale dissipation.
Together with the convergence of the higher-order energy metrics shown in \cref{fig:convergence}, these residuals show that the present simulations close the energy budgets to the accuracy required for the conclusions drawn in the main text.

\section{Spherical-harmonic modal decomposition of drop shape}
\label{app:modal_analysis}

This appendix details the spherical-harmonic projection methodology used to extract the modal decomposition of the drop interface shape from the three-dimensional Basilisk simulation of the $\We\approx11$ prolate water-in-air drop described in \cref{subsec:dropshape_exp_comparison} ($\rho=815.9$, $\Ohd=1.26\times10^{-3}$, $\Oho=6.96\times10^{-4}$).
The key result --- that the axisymmetric ($m=0$) fraction of captured shape energy exceeds $98.6\%$ throughout $t^*\le1.2$ --- is presented in figure~\ref{fig:axi_frac} in \cref{subsec:dropshape_exp_comparison}; here we document the projection formulae and present the supporting diagnostics.

At each saved snapshot, the PLIC reconstruction provides a set of interface facets.
For each facet $f$, let $\bm{x}_f$ be its centroid, $\hat{\bm{n}}_f$ its outward unit normal, and $\mathrm{d}A_f$ its area.
Measuring from the drop centroid $\bm{x}_c$, each facet has a radius $R_f = |\bm{x}_f - \bm{x}_c|$, a colatitude $\theta_f$ from the streamwise axis, and an azimuth $\phi_f$.
The normalised radial deviation is
\begin{equation}
  q_f = \frac{R_f - R_0}{R_0},
  \label{eq:q_app}
\end{equation}
where $R_0$ is the volume-averaged radius.
Each facet also carries a solid-angle weight
\begin{equation}
  \mathrm{d}\Omega_f = \frac{\hat{\bm{n}}_f \cdot \hat{\bm{r}}_f}{R_f^2}\,\mathrm{d}A_f,
  \qquad \hat{\bm{r}}_f = \frac{\bm{x}_f - \bm{x}_c}{R_f},
  \label{eq:domega_app}
\end{equation}
which integrates to $4\pi$ over a closed surface.
The squared spectral amplitude of a spherical harmonic of degree $n$ and order $m$ is then obtained by the self-normalising discrete projection
\begin{equation}
  V_{nm}^\alpha = \frac{\displaystyle\left(\sum_f q_f\,\Phi_{nm}^\alpha(\theta_f,\phi_f)\,\mathrm{d}\Omega_f\right)^2}
                {\displaystyle\sum_f (\Phi_{nm}^\alpha)^2(\theta_f,\phi_f)\,\mathrm{d}\Omega_f},
  \label{eq:Vnm_app}
\end{equation}
where $\alpha\in\{c,s\}$ denotes the cosine configuration $\Phi_{nm}^c = P_n^m(\cos\theta)\cos(m\phi)$ or
the sine configuration $\Phi_{nm}^s = P_n^m(\cos\theta)\sin(m\phi)$.
The two spatial configurations are identical, only rotated by $\pi/(2m)$ about the axis of symmetry.
$P_n^m$ are the associated Legendre polynomials.
Since the denominator is a summation over the same facets and same Rayleigh basis, all normalisations of the solid angle infinitesimals (for axisymmetric simulations) or the Legendre polynomials are cancelled and the ratio is thus independent of the chosen normalisation conventions.
$V_{nm} = V_{nm}^c + V_{nm}^s$, gives the orientation-independent squared amplitude for a specific $(n,m)$ pair.
The squared amplitude and its RMS for a given degree $n$ across all its azimuthal orders are given by
\begin{equation}
  V_n = \sum_{m=0}^{n} V_{nm}, \qquad \tilde{q}_n = \sqrt{V_n}.
  \label{eq:Vn_app}
\end{equation}
$\tilde{q}_n$ is the RMS fractional radial deviation held by degree $n$.
The total shape variance for degree $n$ is thus $q^2_\mathrm{tot} = \tfrac{1}{4\pi}\sum_f q_f^2\,\mathrm{d}\Omega_f$.
This projection is valid only while the drop surface is radially single-valued, i.e., every radially-outward ray from $\bm{x}_c$ pierces the interface exactly once.
Furthermore, given that in principle an infinite number of modes can be excited, the projection is truncated at a finite $n_\mathrm{max}$, which is limited by the spatial resolution of the interface reconstruction and the computational cost of the projection.
The captured fraction $\tilde{q}^2_\mathrm{cap}/q^2_\mathrm{tot}$, where $\tilde{q}^2_\mathrm{cap} = \sum_{n\le n_\mathrm{max}} V_n$, should remain close to unity for a fairly complete decomposition.
The current analysis uses $n_\mathrm{max}=8$, and we will show it to be sufficient for capturing the energetically relevant modes during the pre-bag regime.

The axisymmetric ($m=0$) and non-axisymmetric ($m\ge1$) contributions to the total shape energy are
\begin{equation}
  \tilde{q}^2_\mathrm{axi} = \sum_n V_{n,0}, \qquad
  \tilde{q}^2_\mathrm{non} = \sum_n \sum_{m\ge1} V_{nm},
  \label{eq:vaxi_vnonaxi_app}
\end{equation}
with $\tilde{q}^2_\mathrm{axi} + \tilde{q}^2_\mathrm{non} = \tilde{q}^2_\mathrm{cap}$.

\begin{figure}
\centering
  \begin{subfigure}[t]{0.325\textwidth}
    \insetfig[3,88]{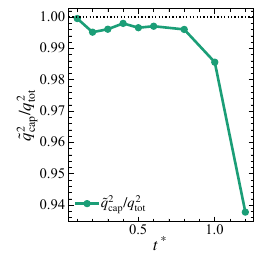}{a}
    \phantomcaption
    \label{fig:mode_analysis-captured_frac}
  \end{subfigure}
  \begin{subfigure}[t]{0.325\textwidth}
    \insetfig[3,88]{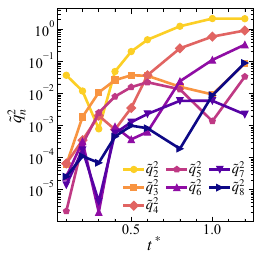}{b}
    \phantomcaption
    \label{fig:mode_analysis-Vn}
  \end{subfigure}
  \par
  \begin{subfigure}[t]{0.325\textwidth}
    \insetfig[3,88]{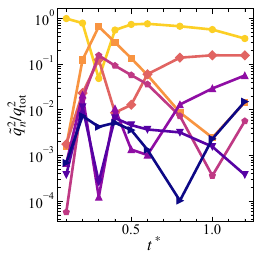}{c}
    \phantomcaption
    \label{fig:mode_analysis-Vn_frac}
  \end{subfigure}
  \caption{\label{fig:mode_analysis}
    Supporting diagnostics for the spherical-harmonic modal decomposition of the $\We\approx11$ 3D simulation (\cref{subsec:dropshape_exp_comparison}).
    (a)~Captured fraction $\tilde{q}^2_\mathrm{cap}/q^2_\mathrm{tot}$: share of total shape variance resolved by the truncated expansion ($n_\mathrm{max}=8$); values near unity confirm that the projection is faithful and the interface remains non-folded.
    (b)~Absolute per-degree variance $\tilde{q}_n^2 = V_n$ for $n=2,3,4$ on a logarithmic scale, showing the growth of modal shape energy as the drop deforms.
    (c)~Fractional per-degree variance $V_n/q^2_\mathrm{tot}$ for $n=2,3,4$: the $n=2$ mode dominates throughout, with higher modes capturing increasing amounts of ambient work as deformation grows toward $t^*=1.2$.
  }
\end{figure}

\Cref{fig:mode_analysis} summarises three supporting diagnostics over the window $t^*\in[0.1,1.2]$.
\Cref{fig:mode_analysis-captured_frac} shows that the captured fraction remains above $93.8\%$ throughout, verifying the sufficiency of the truncated projection ($n_\mathrm{max}=8$).
\Cref{fig:mode_analysis-Vn} shows the absolute per-degree variance $\tilde{q}_n^2 = V_n$ on a logarithmic scale, revealing the growth of modal shape energy as the drop deforms; \cref{fig:mode_analysis-Vn_frac} shows the same quantities normalised by $q^2_\mathrm{tot}$.
An interesting observation is the dominance of even modes over odd modes, with $n=2,4,6$ and $8$ holding the majority of the surface energy by $t^*\approx1.2$.
Even modes have lobes symmetric about the equatorial plane oriented along the axial direction; since both the initial shape and the aerodynamic forcing respect this symmetry, modal energy is preferentially transferred to even modes, while odd modes can only be excited through nonlinear interactions.
Together with the axisymmetric fraction shown in figure~\ref{fig:axi_frac}, these diagnostics confirm that, for normal drop deformation up to $t^*\approx1.2$, modal shape energy is almost exclusively concentrated in axisymmetric ($m=0$) modes, justifying the use of axisymmetric simulations to track the temporal energetics of the drop interface.

\bibliographystyle{jfm}
\bibliography{zotero.bib}

\end{document}